\documentclass[aps,twocolumn]{revtex4}

\usepackage[xdvi]{graphicx}
\usepackage{pdfpages}
\usepackage{latexsym}
\usepackage{amsbsy}
\usepackage{amsmath}
\usepackage{amssymb}
\usepackage{lineno}
\usepackage{multirow}

\begin{document}

\date{\today}

\title{Bridging Quantum and Classical Descriptions of Spin Dynamics in a Dzyaloshinsky–Moriya Trimer}  

\author{Robert Wieser$^{1,2}$, Ra\'ul S\'anchez Gal\'an$^3$}
\affiliation{1. School of Physics and Optoelectronic Engineering,
 Nanjing University of Information Science and Technology, Nanjing
 210044, China \\
2. Jiangsu Key Laboratory for Optoelectronic Detection of Atmosphere and Ocean,
Nanjing University of Information Science and Technology, Nanjing
210044, China \\
3. 4i Intelligent Insights, Tecnoincubadora Marie Curie, PCT Cartuja,
41092 Sevilla, Spain
}

\begin{abstract}
  The spin dynamics of a trimer with Dzyaloshinsky–Moriya (DM) interaction are investigated within a unified Hamiltonian framework that connects quantum-mechanical and semiclassical descriptions. The interpolation between the two regimes is realized by solving the modified Gisin–Schr\"odinger equation, in which the relative weight of quantum coherence and local mean-field contributions is continuously tuned. The resulting dynamical behavior is analyzed and summarized in a ground state diagram that illustrates how the character of the spin motion evolves from fully quantum to semiclassical as the DM interaction is treated at different levels of approximation. In the last part of the publication, the chiral spin dynamics originally proposed by Da-Wei Wang \emph{et al.} is examined theoretically, taking into account its behavior at the boundary between quantum and classical physics.
\end{abstract}

\maketitle

\section{Introduction}
Understanding the transition between quantum and classical spin dynamics remains one of the central challenges in the description of magnetic systems at the nanoscale. While the fully quantum-mechanical treatment captures coherence, entanglement, and tunneling effects, semiclassical approaches provide valuable insights into collective motion and macroscopic behavior. Bridging the quantum and classical regimes is essential for a unified understanding of spin dynamics, and a variety of approaches addressing this crossover have been developed in the literature. On a conceptual level, the emergence of classical magnetic order from quantum states has been discussed within the framework of decoherence and environment-induced superselection (“einselection”), where robust pointer states corresponding to classical spin configurations are selected through environmental coupling \cite{zurekRMP03,donkerPRB16}.

Complementary to this viewpoint, recent works have demonstrated that classical magnetic textures (such as spatial arrangements of spins) can be reconstructed from finite-size quantum spectra using spectral signatures such as the Anderson tower of states (a series of low-energy quantum states that signal emerging magnetic order in finite systems), allowing, for instance, the reconstruction of classical textures like Skyrmions from finite-size quantum spectra \cite{sotnikovPRX23}. 

In addition, direct dynamical comparisons between the Schrödinger equation and semiclassical Landau–Lifshitz-Gilbert dynamics have revealed a remarkable quantitative agreement even for spin-1/2 systems, highlighting the robustness of semiclassical descriptions beyond their naive range of validity \cite{schubertPRB21},  see also related open-system and hybrid treatments in \cite{heitmannPRR22}.

In contrast to these approaches, the present work focuses on a dynamical hybrid framework that explicitly interpolates between fully quantum and local mean-field descriptions at the level of the equations of motion. This allows us to study the continuous crossover between quantum-entangled and increasingly classical spin states within a single, unified model. 
In this work, we investigate such a crossover within a trimer model including Dzyaloshinsky–Moriya interaction (DMI), which naturally introduces chiral spin textures and noncollinear configurations. The approach connects quantum and semiclassical descriptions within a single Hamiltonian framework, allowing the continuous interpolation between both limits and providing a basis to analyze how chiral spin dynamics emerge across the quantum–classical boundary.
The spin system which shall be considered with this theory is a trimer with Dzyaloshinsky-Moriya interaction (DMI). The study of magnetic nanostructures with chiral order has gained considerable importance in recent years. In particular, the DMI, which arises from spin-orbit coupling in systems lacking inversion symmetry, fosters the formation of noncollinear and topologically intriguing spin arrangements, such as vortices \cite{wieserPRB06,lebeckiPRB14,bhattacharjeeJPCM25,vigoJAP25,khoslaPRB25} and skyrmions \cite{everschorJAP18,sieglPRR22,schickPRR24,petrovicRMP25,wuPQE25,vedmedenkoPRAppl25,komeljPRR25}. In this context, spin clusters represent a particularly interesting model system. They allow the investigation of complex magnetic phenomena—including chirality, entanglement, and coherence within minimal geometries. Furthermore, they provide an ideal platform for analyzing the transition between quantum and classical behavior, since even three interacting spins exhibit a multitude of quantum mechanical correlations and collective effects. A central idea is to control the magnetic structure or quantum state by tuning the interactions or applying external perturbations \cite{parkinsonJPCM01,marchukovNATURECOMM16,blokhinaIEEEAccess20,wangIEEEQE23,bateyFICT15,hanzeSCIADV21,wieserAP25}. Spin clusters serve here as prototypical systems, representing not only magnetic nanostructures on surfaces or within molecules, but also effective spin states in quantum dots \cite{burkardFP00}, photonic lattices \cite{slussarenkoAPR19}, or ultracold atomic ensembles \cite{lanyonScience11}.  Previous works -- for example, by S. Loth \emph{et al.} \cite{lothSCIENCE12,delgadoEPL15,yanSCIADV17}, or R. Wiesendanger \emph{et al.} \cite{khajetooriansSCIENCE11,khajetooriansNATPHYS12,beckPRB23,rodriguezACSNANO24}, have investigated magnetic clusters on surfaces. These experiments showed that these clusters are suitable as logic elements for quantum computing or information storage. The use of spin-polarized scanning tunneling microscopes (SP-STM) is also suitable for the precise excitation of individual spins. As an alternative to local excitations, e.g., with scanning tunneling microscope -- electron spin resonance (STM-ESR) \cite{willkeSCIADV18,astPRR24,choiNA25}, individual spins can also be excited with global excitations, although different chemical environments are required \cite{singhPRA16,singhPRA18,jonesPNMRS24}. A. Stepanov \emph{et al.} proposed Heisenberg-exchange-free nano-skyrmions and investigated their properties, including their potential use as robust and compact data storage elements \cite{stepanovJPCM19}. Moreover, such nano-skyrmions have been suggested as candidates for implementing skyrmion-based qubits at the nanoscale \cite{lohaniPRX19,gauyacqJPCM19,sotnikovPRB21,psaroudakiPRL21,hallerPRR22}. Da-Wei Wang \emph{et al.} have experimentally realized Heisenberg-exchange-free spin clusters \cite{wangNATUREPhys19}, where the interaction is governed by the Dzyaloshinsky-Moriya interaction, as proposed by E. A. Stepanov \emph{et al.}. The smallest such spin cluster is a spin trimer, which consists of three spins on a circle and serves as a fundamental building block for larger spin configurations. The authors demonstrated the existence of chiral spin dynamics induced by the Dzyaloshinsky-Moriya interaction, using superconducting spin circuits to realize and probe the time-dependent behavior of these systems. An analogous experiment using ultracold atoms with a photon-mediated Dzyaloshinsky-Moriya interaction instead of superconducting circuits was carried out later by Shuyue Wang \emph{et al.} \cite{wangPRA24}. The spin dynamics investigated by Da-Wei Wang \emph{et al.} and Shuyue Wang \emph{et al.} is a purely quantum mechanical phenomenon with no classical analogue. This raises fundamental questions: What happens when a quantum spin cluster is approximated using classical models? How does the quantum-to-classical transition manifest, and what changes occur in the internal structure and correlations of the cluster? This work aims to analytically investigate the dynamics of a spin trimer confined to the $XY$-plane with a $z$-oriented Dzyaloshinsky-Moriya interaction and an external magnetic field along the $z$-axis. A particular focus is on understanding the emergent chiral structures and the behavior of the system during the transition from a purely quantum mechanical regime to one that can be described classically. Thus, this work contributes to the fundamental understanding of quantum coherence, entanglement, and chirality in minimal spin structures.

The publication is organized as follows: The next section introduces the underlying model and the computational method used to study the spin system. Section III discusses the ground state diagram resulting from the quantum-classical transition. Section IV deals with the chiral spin dynamics originally proposed by Da-Wei Wang \emph{et al.}. In this section, the analytical expressions of spin dynamics are presented and discussed. The publication concludes with a summary in Section V.

\section{Theoretical Model}\label{sect_theo}
We consider a system of three spins \(S = 1/2\) (a \emph{spin trimer}) with periodic bonds $\vec{S}_1\leftrightarrow \vec{S}_2$, $\vec{S}_2 \leftrightarrow \vec{S}_3$, $\vec{S}_3 \leftrightarrow \vec{S}_1$, described by the Hamilton operator
\begin{equation} \label{HamTotal}
  \hat{\mathrm{H}} = \hat{\mathrm{H}}_B + \hat{\mathrm{H}}_D \;.
\end{equation}
The two contributions are defined as
\begin{subequations}
\begin{align}
  \hat{\mathrm{H}}_B &= -B \sum_{n} \hat{S}_n^z \;, \label{Zeeman} \\
  \hat{\mathrm{H}}_D &= \sum_{\langle n,m \rangle} \Big[ \vec{D} \cdot \left( \vec{S}_n \times \vec{S}_m \right) 
    + \vec{D}_{\mathrm{LMF}} \cdot \left( \vec{S}_n \times \langle \psi | \vec{S}_m | \psi \rangle \right) \Big]\;,
\end{align}
\end{subequations}
with $\vec{S}_n = (\hat{S}_n^x, \hat{S}_n^y, \hat{S}_n^z)^{\mathrm{T}}$ the spin operator at lattice site $n$. 

The term $\hat{\mathrm{H}}_B$ represents the Zeeman coupling between the spins and an external magnetic field of strength $B$, applied along the $z$-axis. $B = \mu_S B_{\mathrm{phys}}$, where $B_{\mathrm{phys}}$ is the physical
magnetic field and $\mu_S = \gamma \hbar/2$, with $\gamma$ the gyromagnetic ratio, the magnetic moment.
Thus, $B$ carries units of energy.

The term $\hat{\mathrm{H}}_D$ describes the Dzyaloshinsky-Moriya (DM) interaction between nearest-neighbor spins, denoted by the $\langle n,m\rangle$ symbol in the sum. $\hat{\mathrm{H}}_D$ includes both quantum and semiclassical contributions. The DM interaction is characterized by the vectors $\vec{D}$ and $\vec{D}_{\mathrm{LMF}}$, both assumed to be oriented along the $z$-axis, with magnitudes $D$ and $D_{\mathrm{LMF}}$, respectively. 

The first part of $\hat{\mathrm{H}}_D$ corresponds to the fully quantum Dzyaloshinsky-Moriya interaction involving two spin operators. It captures genuine quantum correlations, such as fluctuations and entanglement between spins.
The second part employs a local mean-field (LMF) approximation, in which each spin interacts only with the expectation value $\langle \psi | \vec{S}_m | \psi \rangle$ of the other spins, with the quantum state $|\psi \rangle$ determined by solving the modified Gisin-Schr\"odinger equation, Eq.~(\ref{eqGisin}). The detailed calculation is described later in the text. This approximation neglects all quantum correlations, corresponding to a purely classical description of the DM interaction. To keep the total interaction strength constant, we impose
\begin{equation}
    D_{\mathrm{LMF}} = 1-D .
    \label{Dnorm}
\end{equation}
This convention ensures that all parameters $B$, $D$, and $D_{\mathrm{LMF}}$ share the same units of energy.

By continuously varying $D$ from one to zero, we interpolate between the fully quantum limit ($D = 1$, $D_{\mathrm{LMF}} = 0$) and the fully classical limit ($D = 0$, $D_{\mathrm{LMF}} = 1$). This interpolation keeps the overall interaction strength constant while allowing for a gradual reduction of quantum correlations. The idea is to weight the fully quantum-mechanical description and the semi-classical description using local mean fields, while keeping the total interaction strength fixed, thereby allowing for a continuous crossover between quantum-correlated and mean-field-dominated regimes. Although a microscopic derivation of this normalization is not yet available, our approach can be regarded as a phenomenological ansatz that mimics the progressive decoherence and coarse-graining of spin correlations. This parameterization finds its physical justification in the theory of open quantum systems. It is well established that the interaction of a quantum system with its environment leads to decoherence, a process that suppresses quantum superpositions and selects robust 'classical' states (often referred to as pointer states) \cite{breuerBOOK07, zurekRMP03}. In the specific case of magnetic clusters, Donker \emph{et al.} have demonstrated that environmental coupling drives the system toward classical magnetic configurations, such as the N\'eel state, effectively washing out quantum entanglement \cite{donkerPRB16}.

Consequently, our hybrid model can be interpreted as a phenomenological description of this decoherence process. The fully quantum DMI ($D$) generates entanglement, while the Local Mean-Field contribution ($D_{\mathrm{LMF}}$) enforces a product-state description typical of the classical limit. By continuously tuning the weights such that $D+D_{\mathrm{LMF}}=1$, we effectively model the renormalization of the coherent coupling strength due to environmental averaging, allowing for a continuous crossover from an entangled quantum trimer to a semiclassical spin texture.

To determine the ground states of the Hamiltonian and analyze the spin dynamics, we solve the modified Gisin-Schr\"odinger equation \cite{gisinHelvPhysActa81,gisinPhysica82,wieserEPJB15,wieserJPComm19},
\begin{equation}
    i\hbar (1 + \alpha^2) \frac{d}{dt} |\psi\rangle
    = \hat{\mathrm{H}} |\psi\rangle
      - i\hbar \alpha
      \big(
        \hat{\mathrm{H}} - \langle \psi | \hat{\mathrm{H}} | \psi \rangle
      \big) |\psi\rangle .
    \label{eqGisin}
\end{equation}
The modified Gisin equation adds a relaxation term proportional to the damping constant $\alpha$ to the time-dependent Schr\"odinger equation.
This term preserves the norm of $|\psi\rangle$ while driving non-eigenstates toward stationary eigenstates of $\hat{\mathrm{H}}$, effectively modeling decoherence or energy relaxation.

The solution of Eq.~(\ref{eqGisin}) is self-consistent because the quantum state $|\psi\rangle$ is needed to evaluate the spin expectation values $\langle \psi | \vec{S}_m | \psi \rangle$, which in turn enter the Hamiltonian $\hat{\mathrm{H}}$. Starting from an initial state $|\psi(t)\rangle$, either random or close to the expected ground state, the system evolves toward a stationary state after several self-consistent iterations. The iteration process is described as follows: The initial quantum state $|\psi(t)\rangle$ at time $t$ is used to determine the Hamiltonian $\hat{\mathrm{H}}$ and energy expectation value $E(t) = \langle \psi(t) | \hat{\mathrm{H}} | \psi(t) \rangle$ and to solve the modified Gisin-Schr\"odinger equation leading to a new quantum state $|\psi(t+\mathrm{d}t)\rangle$, which initiates the new iteration cycle. The iteration stops when $\mathrm{d}E(t) = E(t+\mathrm{d}t)-E(t) \approx 0$. The iteration leads to a monotonically decreasing energy and the quantum state $|\psi\rangle$ becomes an eigenstate of $\hat{\mathrm{H}}$.   
With a suitable choice of the initial state, the resulting stationary state represents the ground state of the spin trimer, from which the ground state diagram can be constructed.

It should be emphasized that the modified Gisin-Schr\"odinger equation represents the \emph{quantum-mechanical analog} of the Landau-Lifshitz-Gilbert (LLG) equation in the case of the pure local mean-field limit ($D = 0$, $D_{\mathrm{LMF}} = 1$). In this scenario,
Eq.~(\ref{eqGisin}) reduces to a single-spin dynamics governed by the classical LLG equation. Moreover, the local mean field can be interpreted as the effective magnetic field
\begin{equation}
    \vec{B}^{\mathrm{eff}}_n
    = -\frac{1}{\mu_S}
      \frac{\partial {\cal H}}{\partial \vec{\cal S}_n} .
    \label{eqBeff}
\end{equation}
For details please see the appendix. In the classical description, each spin interacts with the others solely through this effective field, resulting in a single-spin precession subject to damping. 
Here, ${\cal H}$ denotes the classical Hamilton function, $\vec{\cal S}_n$ the classical spin, and $\mu_S = |\vec{\cal S}_n|$ the magnetic moment of the classical spin. Throughout this paper, we set $\hbar = 1$.

\section{Description of the trimer using a quantum to classical spin dynamics bridging}
\begin{figure}[ht]
  \begin{center}
    \includegraphics*[width=7.0cm,bb = 70 350 515 770]{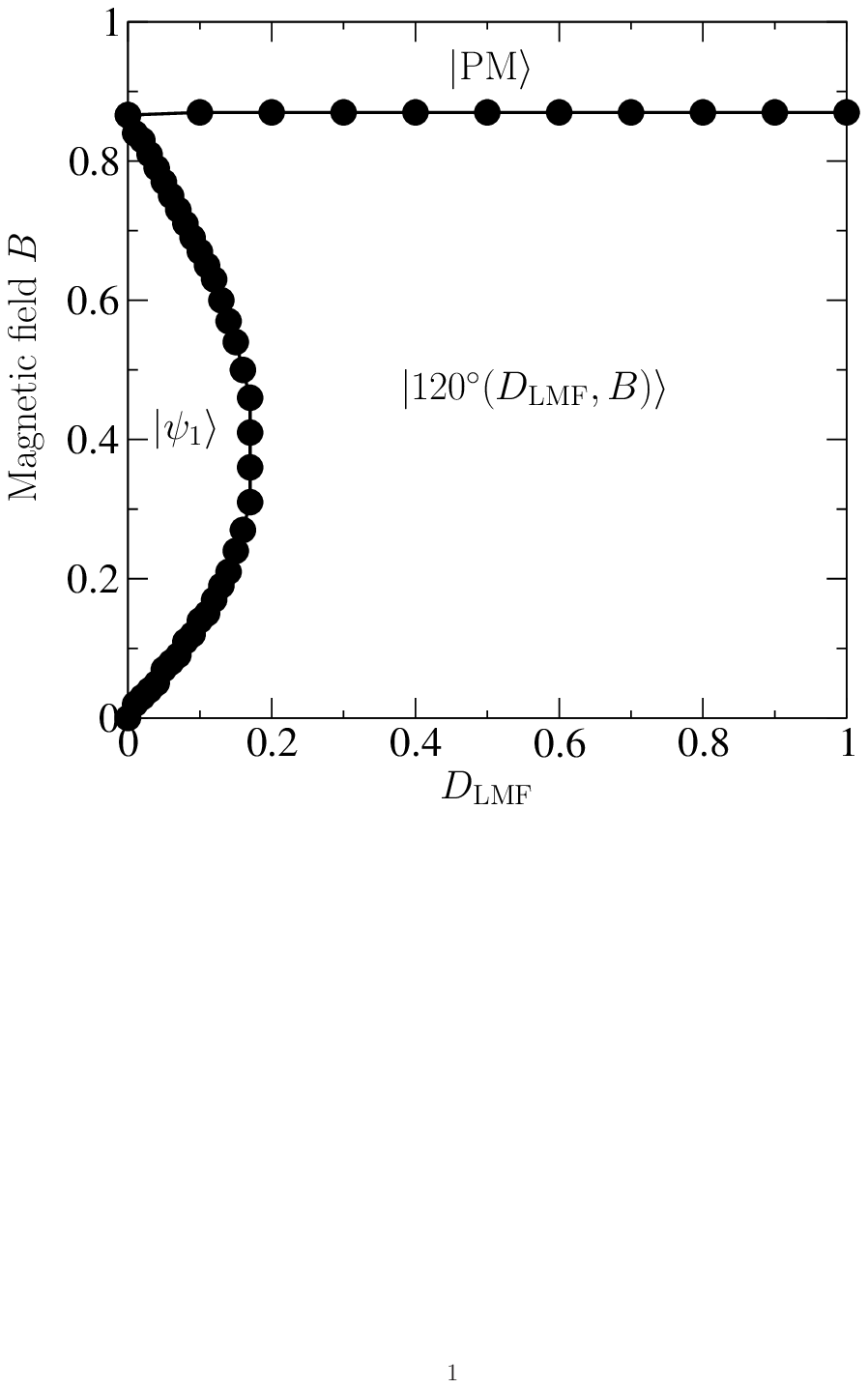}
  \end{center}
  \caption{Ground state diagram: magnetic field $B$ over the local mean-field contribution to the Dzyaloshinsky-Moriya interaction $D_{\mathrm{LMF}}$. The dotted lines represent the calculated boundaries of the different quantum states. Details regarding the quantum states and crossover can be found in the text.}  
  \label{f:pic1}
\end{figure}

The ground state diagram of the trimer under consideration is depicted in Fig.~\ref{f:pic1}. The transition parameter $D_{\mathrm{LMF}}$ quantifies the degree of classicality of the spin system: $D_{\mathrm{LMF}} = 0$ corresponds to a purely quantum mechanical system, while $D_{\mathrm{LMF}} = 1$ corresponds to a semiclassical system. 
It should be emphasized that all states remain quantum. The term ``semiclassical'' here refers to the local mean-field approximation, which yields
product coherent states without entanglement. In this limit the description coincides with classical spin configurations.
The ordinate of the graph represents the influence of the external magnetic field $B$. The ground state diagram 
shows three different types of ground states $|\psi_1\rangle$, $|120^\circ(D_{\mathrm{LMF}},B)\rangle$, and $|\mathrm{PM}\rangle$, which we will now describe. The simplest ground state to describe is $|\mathrm{PM}\rangle$. This quantum state is a product state with all three spins aligned in the direction of the magnetic field, $|\mathrm{PM}\rangle = |\!\uparrow\uparrow\uparrow\rangle$. One could call this state ferromagnetic, but in this case, there is no exchange coupling, and the Dzyaloshinsky-Moriya interaction has no effect due to the parallel alignment of the spins. It is much more of a paramagnetic state. The quantum state $|\mathrm{PM}\rangle$ occurs when the magnetic field strength $B$ exceeds the critical value $B_C = \sqrt{3}/2$. This value depends on the strength of the Dzyaloshinsky-Moriya interaction, but this threshold does not depend on whether the description is quantum mechanical or semiclassical. For $B > B_C$, there are no quantum fluctuations, so the spin expectation values are $|\langle \mathrm{PM}|\vec{S}_n|\mathrm{PM}\rangle| = 1/2$. Furthermore, the energy of this quantum state is determined solely by the applied magnetic field.

The quantum state 
\begin{eqnarray}
  |\psi_1\rangle = \frac{1}{\sqrt{3}}\left(e^{-\frac{2i\pi}{3}}|\!\downarrow\uparrow\uparrow \rangle + |\!\uparrow\downarrow\uparrow \rangle + e^{+\frac{2i\pi}{3}}|\!\uparrow\uparrow\downarrow \rangle \right) \;,
\end{eqnarray}
is an eigenstate of the Hamilton operator $\hat{\mathrm{H}}$, and can be easily calculated by exact diagonalization in the case $D_{\mathrm{LMF}} = 0$. The state $|\psi_1\rangle$ is a modified $W$-state with cyclic phases ($C_3$-symmetry) caused by the Dzyaloshinsky-Moriya interaction. Although the spin expectation values $\langle \psi_1 |\vec{S}_n|\psi_1\rangle$, have a parallel alignment, this quantum state exhibits chirality, defined for a quantum state $|\psi\rangle$ as,
\begin{eqnarray} \label{ChiralityDef}
    \chi = \langle \psi |\vec{S}_1\cdot(\vec{S}_2 \times \vec{S}_3)|\psi\rangle\;.
\end{eqnarray}
However, there are reduced magnetizations of $|\langle \psi_1 |\vec{S}_n|\psi_1\rangle| = 1/6$, which indicates entanglement. The entanglement can be quantified using the von Neumann entropy ${\cal S}_n$. In this case, the three spins have von Neumann entropy ${\cal S}_n \approx 0.918296$, indicating that $|\psi_1\rangle$ is highly, but not entirely, entangled. Chirality can be quantified as $\chi \approx 0.433013$. This value is higher than the classical chirality value $\chi = 1/8$ for a principal angle of $90^\circ$ and $S = 1/2$. The quantum state remains unchanged with an increasing magnetic field and is also stable with respect to a small $D_{\mathrm{LMF}}$.

In the absence of a magnetic field, $B = 0$, the state $|\psi_1\rangle$ is degenerate, and together with the eigenstate \(
|\psi_2\rangle
= \frac{1}{\sqrt{3}}(
    e^{-2\pi i/3}\,|\downarrow\downarrow\uparrow\rangle
   + |\downarrow\uparrow\downarrow\rangle
   +  e^{2\pi i/3}\,|\uparrow\downarrow\downarrow\rangle),
\) forms a degenerate doublet at $B=0$ in the quantum limit ($D_{\mathrm{LMF}}=0$). The superposition 
\begin{eqnarray}
  |\psi_{12}\rangle = \frac{1}{\sqrt{2}}\Big(|\psi_1\rangle + |\psi_2\rangle \Big)\;,
\end{eqnarray}
then results in another ground state, which is characterized by the fact that the magnetic moments (spin expectation values $\langle \psi_{12} | \vec{S}_n |\psi_{12} \rangle$) are localized in the $XY$-plane and have an angle of $120^\circ$ to each other. The magnetizations in this case are $|\langle \psi_{12} |\vec{S}_n|\psi_{12} \rangle| = 1/3$, indicating a reduction in entanglement. The value of the von Neumann entropies are ${\cal S}_n \approx 0.650022$. 

The quantum states $|120^\circ(D_{\mathrm{LMF}},B) \rangle$ occupy most of the ground state diagram. These quantum states are characterized by the spin expectation values $\langle 120^\circ(D_{\mathrm{LMF}},B) |\vec{S}_n|120^\circ(D_{\mathrm{LMF}},B) \rangle$, which form the three segments joining the vertices with the centroid of an equilateral triangle. This structure arises from the geometric frustration caused by the periodic boundary conditions. The Dzyaloshinsky–Moriya interaction attains its minimum energy when the spins are oriented at $90^\circ$ to each other; however, this configuration is not possible in the cyclic geometry of the trimer. The resulting behavior is analogous to that of a trimer with antiferromagnetic exchange interaction in a ring structure. For a non-zero external magnetic field ($B \neq 0$), this quantum state develops a non-zero magnetization component $\langle \hat{S}_n^z \rangle$ along the $z$-axis. When $B \geq B_C$, the spin-flip transition is complete and the three spins are aligned in parallel. In the semiclassical limit, $D_{\mathrm{LMF}} = 1$, these quantum states can be expressed as a product of coherent spin states \cite{radcliffeJPA71}:
  \begin{eqnarray} \label{ThetaPhi}
    |120^\circ(D_{\mathrm{LMF}} = 1,B)\rangle = \bigotimes_{n=1}^3 |\theta_n \phi_n\rangle \;, 
  \end{eqnarray}  
  where
  \begin{align} 
    |\theta_n \phi_n\rangle &= \left(1+|\mu_n|^2\right)^{-S} \exp(\mu_n \hat{S}_-)|0\rangle \;, \nonumber \\
    \mu_n &= \tan\!\left(\frac{\theta_n}{2}\right) \exp(i\phi_n) \;, \nonumber 
  \end{align}  
with $|0\rangle$ denoting the eigenstate satisfying $\hat{S}_z|0\rangle = S|0\rangle$, with $S$ being the spin quantum number and $\hat{S}_- = \hat{S}_x - i \hat{S}_y$ the lowering operator. Here, $\theta_n$ and $\phi_n$ are the polar and azimuthal angles, respectively, that describe the spatial orientation of the $n$-th spin ($n = 1,2,3$). The values of $\theta_n$ and $\phi_n$ depend on the strength of the magnetic field $B$ as well as on the Dzyaloshinsky-Moriya interaction coefficients $D$ and $D_{\mathrm{LMF}}$.

In the $S = 1/2$ limit the coherent spin states (\ref{ThetaPhi}) are identical to the qubit states:
\begin{eqnarray}
  |\theta_n \phi_n\rangle = \cos(\theta_n/2)|\!\uparrow\rangle + e^{i\phi_n}\sin(\theta_n/2)|\!\downarrow\rangle \;.
\end{eqnarray}

While the ground states $|\psi_1\rangle$ and $|\mathrm{PM}\rangle$ are entirely independent of both $B$ and $D_{\mathrm{LMF}}$, the states $|120^\circ(D_{\mathrm{LMF}},B)\rangle$ depend strongly on the magnetic field strength $B$ and on the degree of classicality $D_{\mathrm{LMF}}$. Increasing either $B$ or $D_{\mathrm{LMF}}$ reduces the entanglement of these states while increasing their magnetization, $|\langle 120^\circ(D_{\mathrm{LMF}},B) |\vec{S}_n|120^\circ(D_{\mathrm{LMF}},B) \rangle|$. The change in magnetization with magnetic field observed here is consistent with the results of Vogt and Kettemann~\cite{vogtAnnPhys09} on quantum fluctuations in spin systems.

\begin{figure}[t]
  \begin{center}
    \includegraphics*[width=4.0cm,bb = 70 450 510 770]{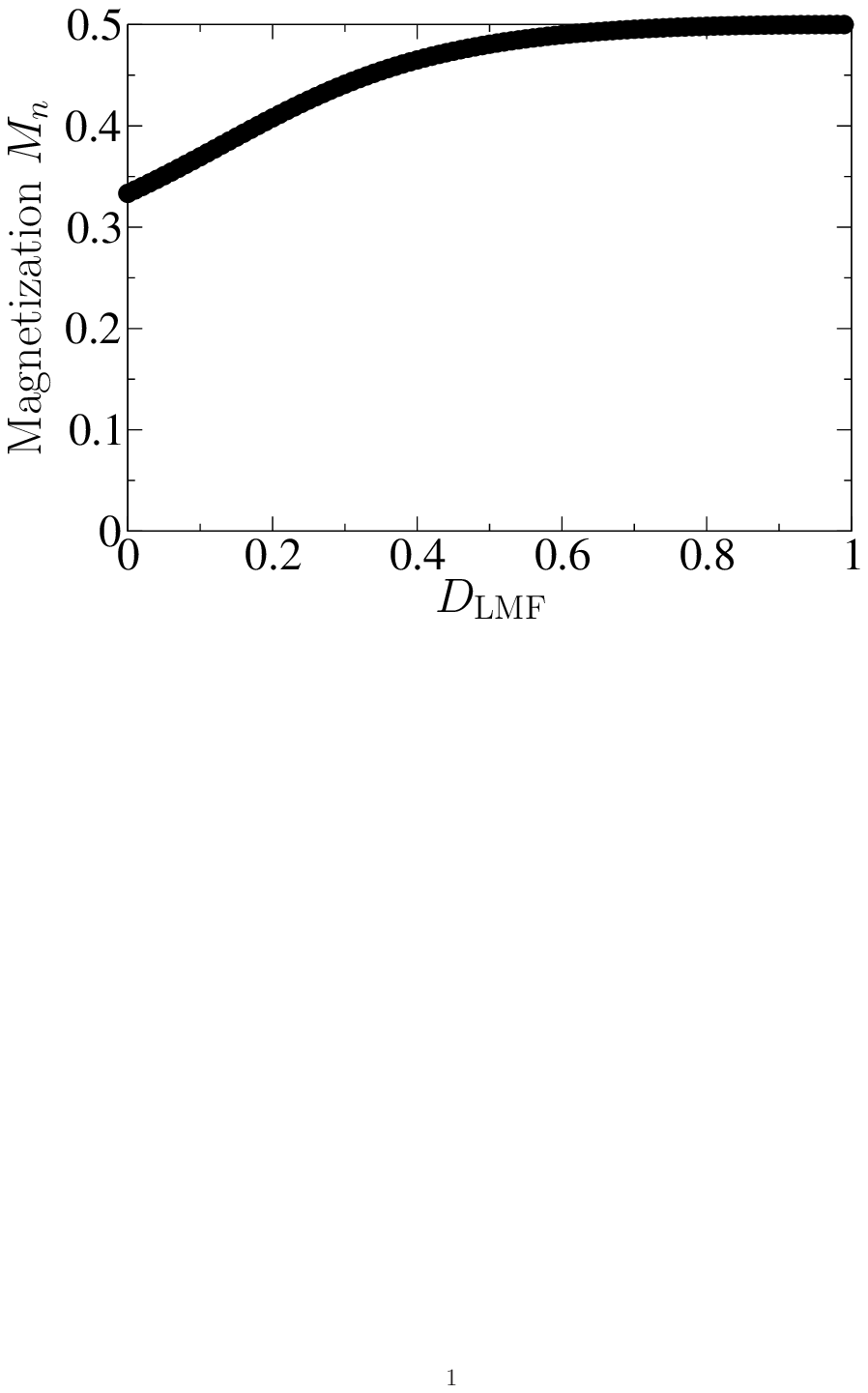}
    \includegraphics*[width=4.0cm,bb = 70 450 510 770]{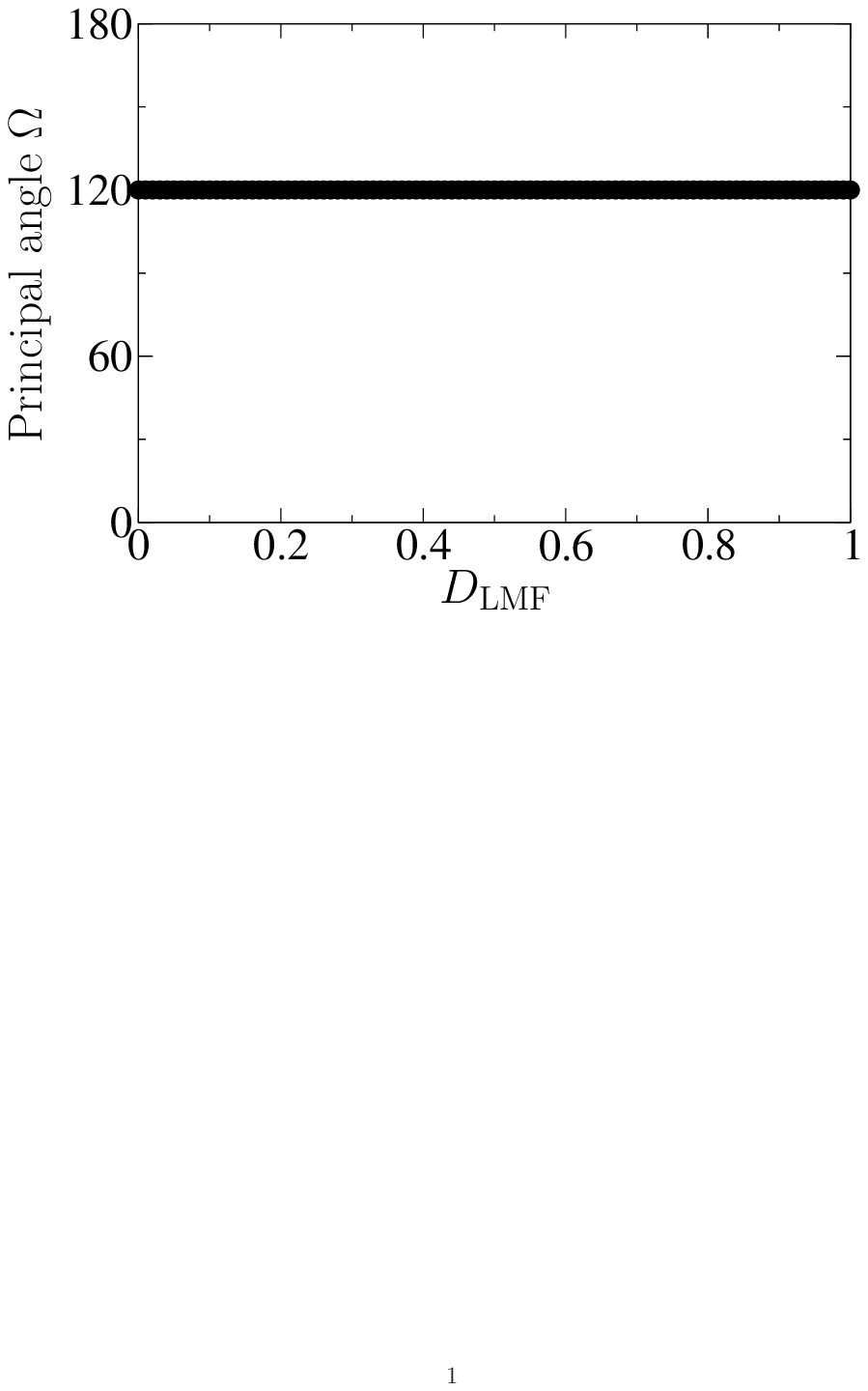}
    \includegraphics*[width=4.0cm,bb = 70 450 510 770]{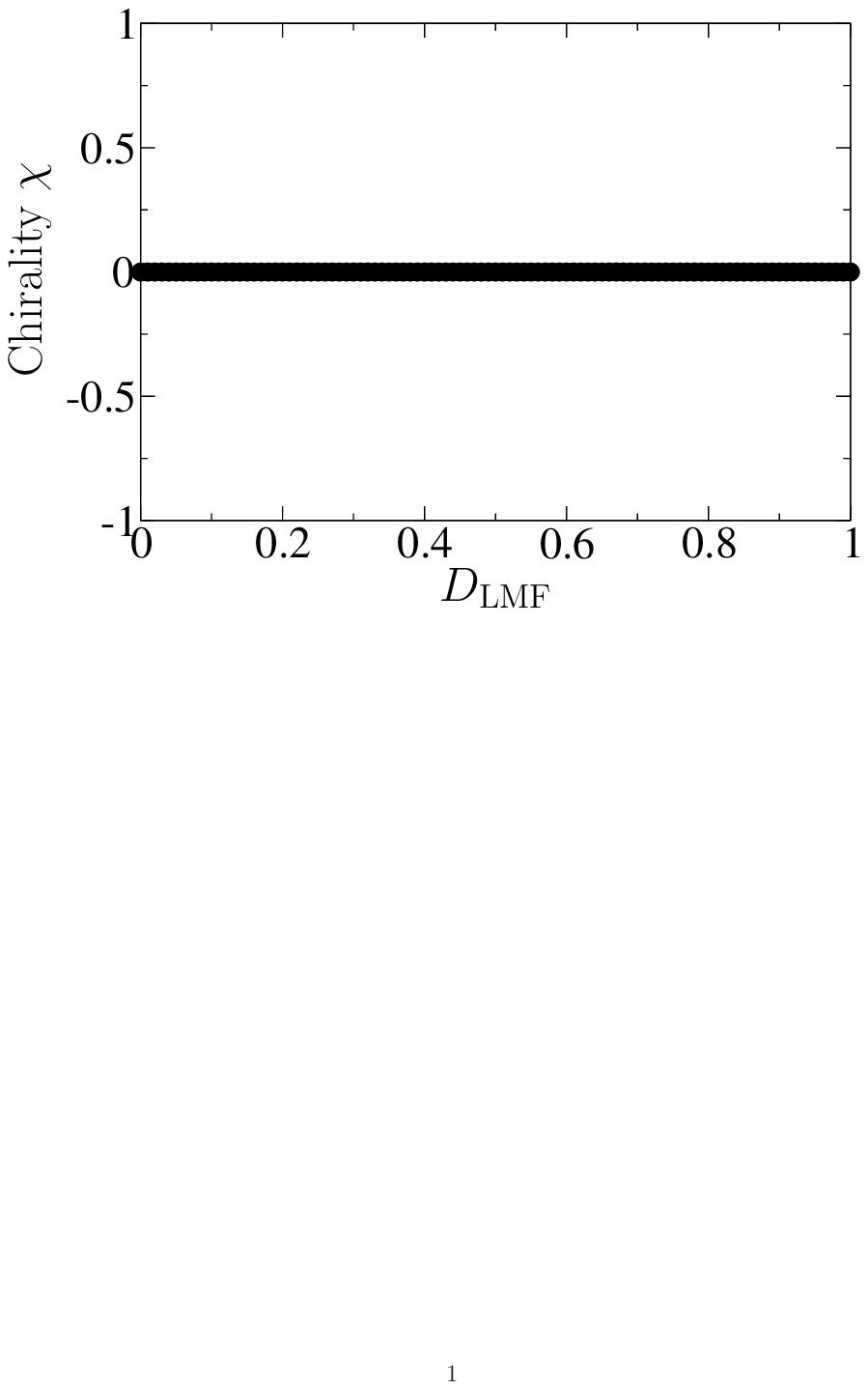}
    \includegraphics*[width=4.0cm,bb = 70 450 510 770]{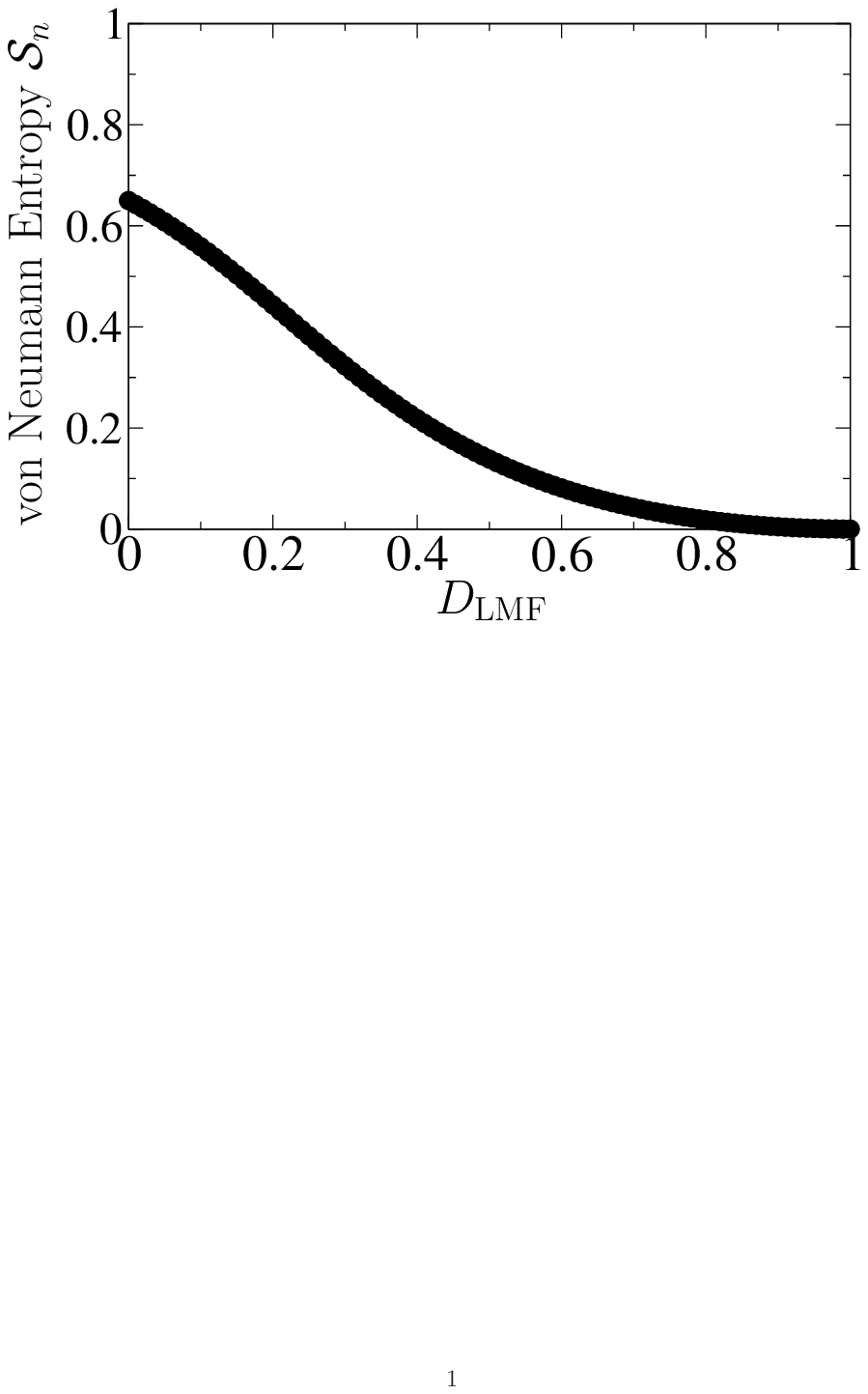}
  \end{center}
  \caption{Magnetization, principal angle, chirality, and von Neumann entropy (as a measure of entanglement) as functions of the classicality parameter $D_{\mathrm{LMF}}$ in the absence of an external magnetic field, $B = 0$.}  
  \label{f:pic2}
\end{figure}
\begin{figure}[t]
  \begin{center}
    \includegraphics*[width=4.0cm,bb = 70 450 510 770]{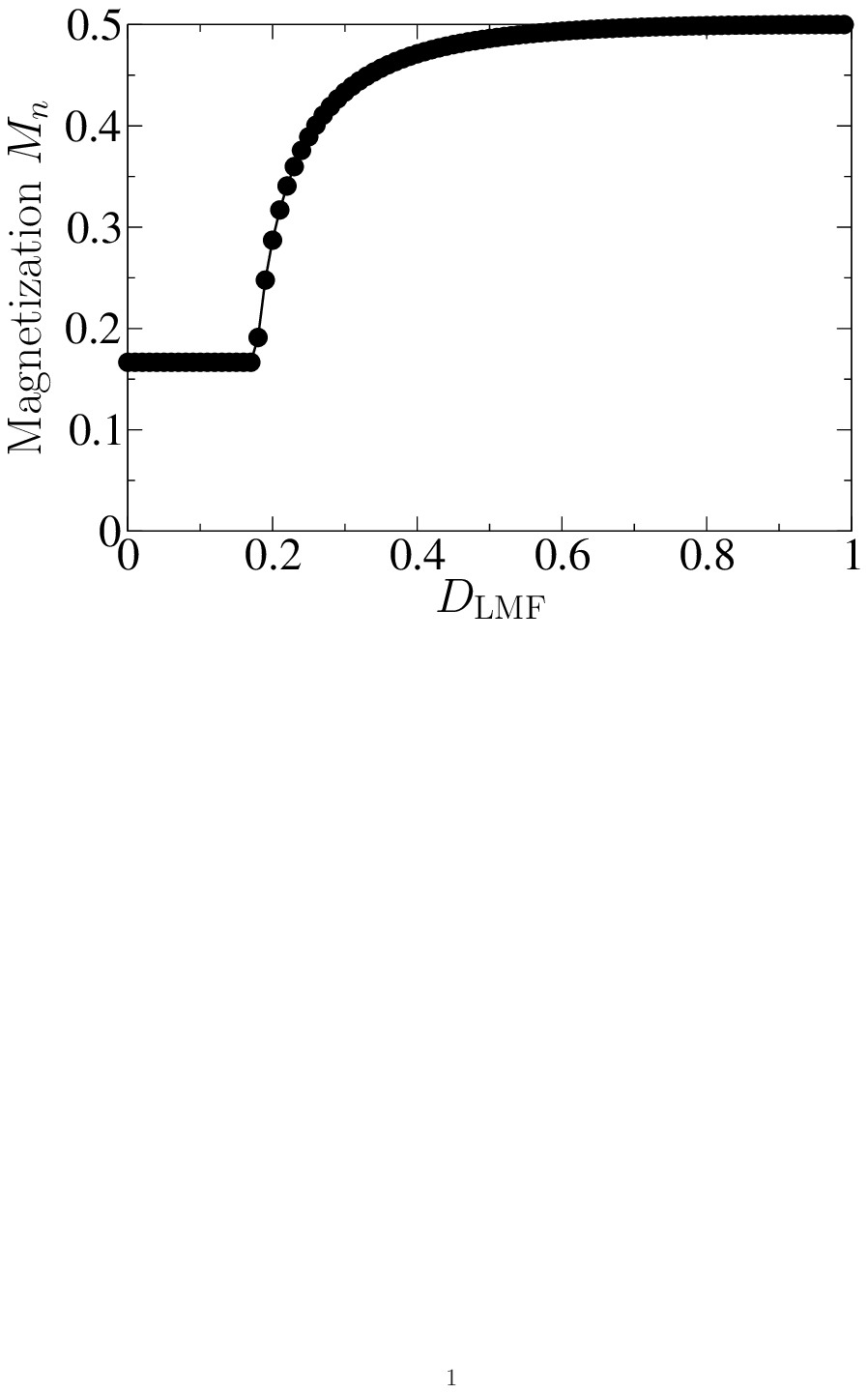}
    \includegraphics*[width=4.0cm,bb = 70 450 510 770]{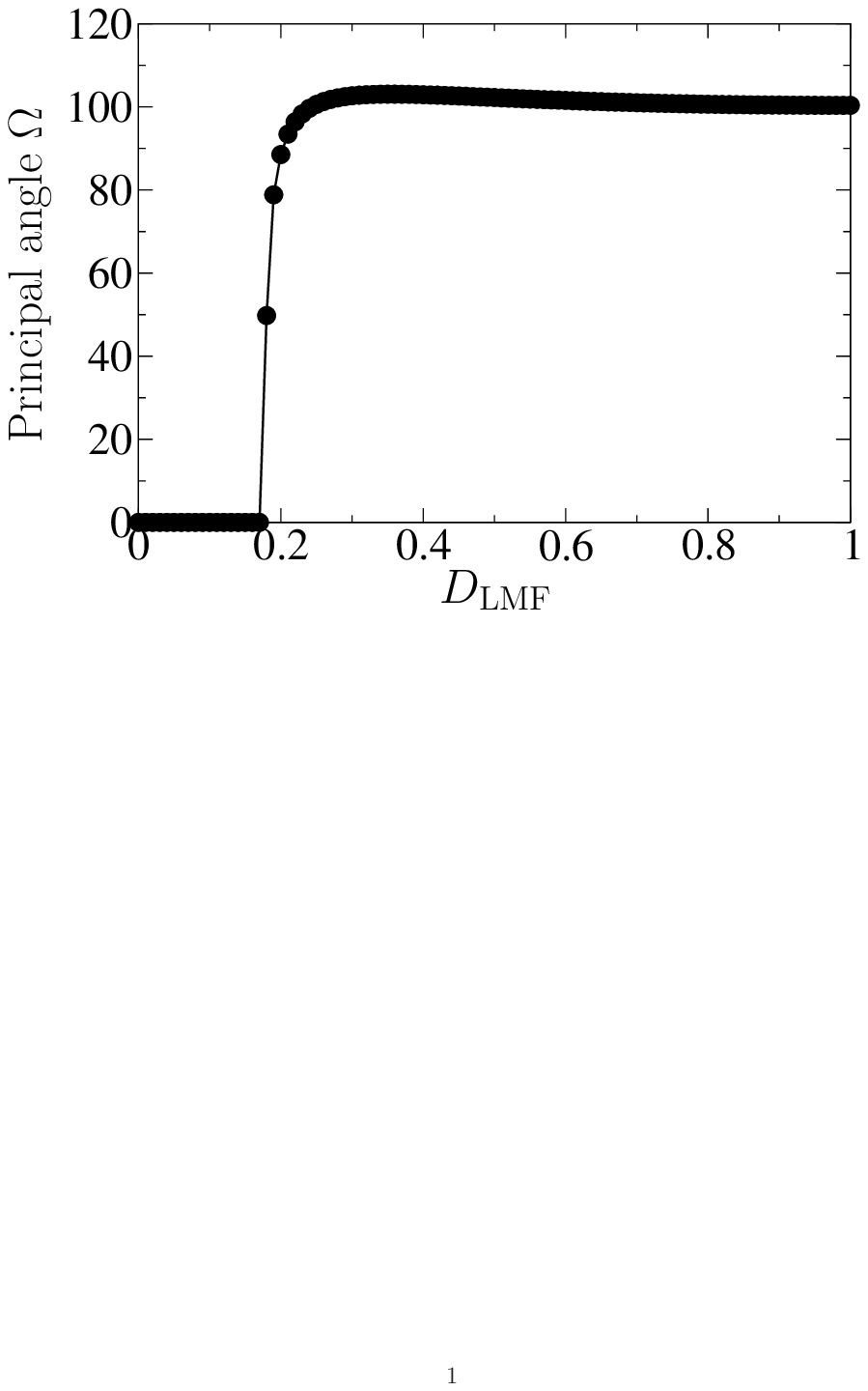}
    \includegraphics*[width=4.0cm,bb = 70 450 510 770]{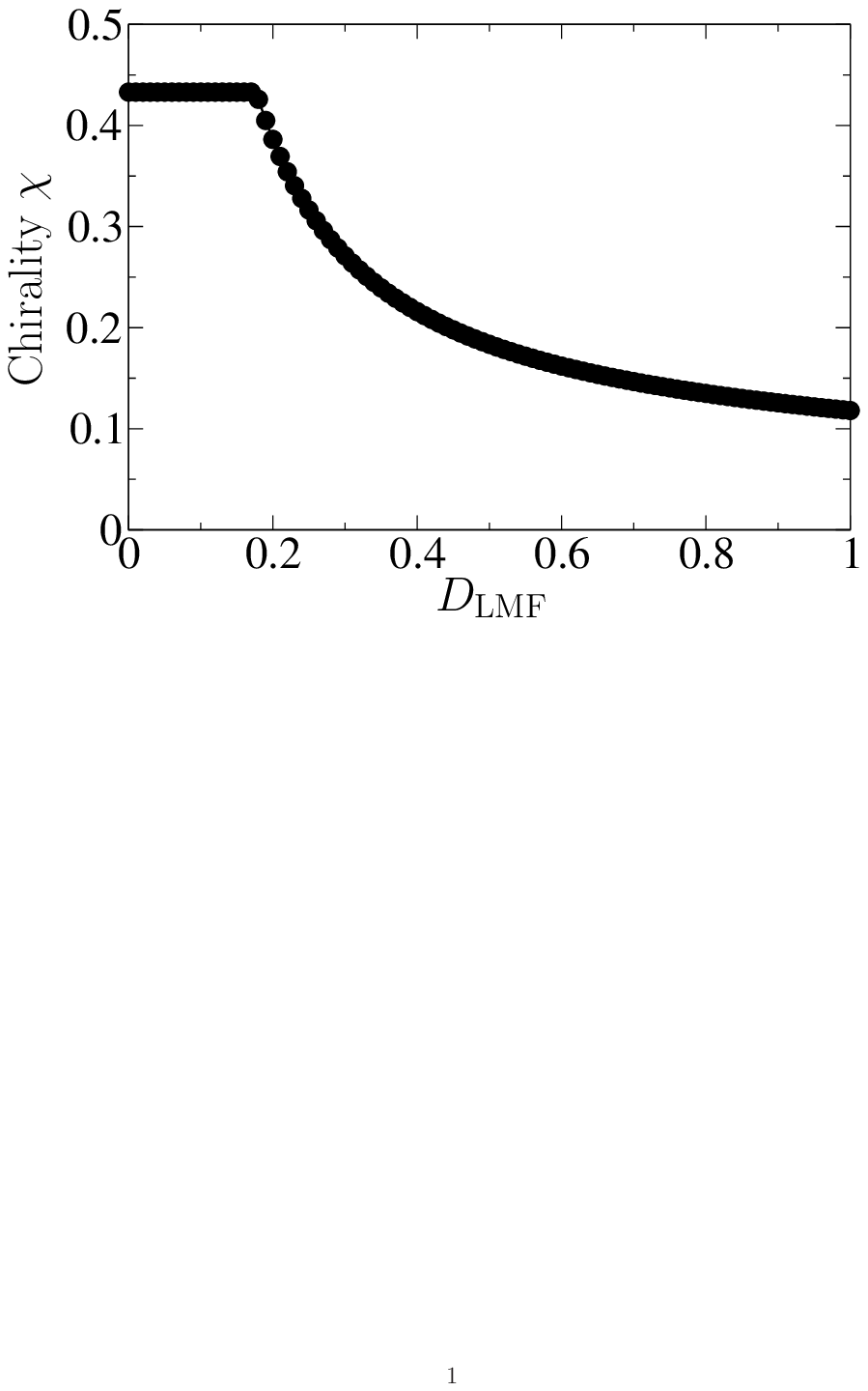}
    \includegraphics*[width=4.0cm,bb = 70 450 510 770]{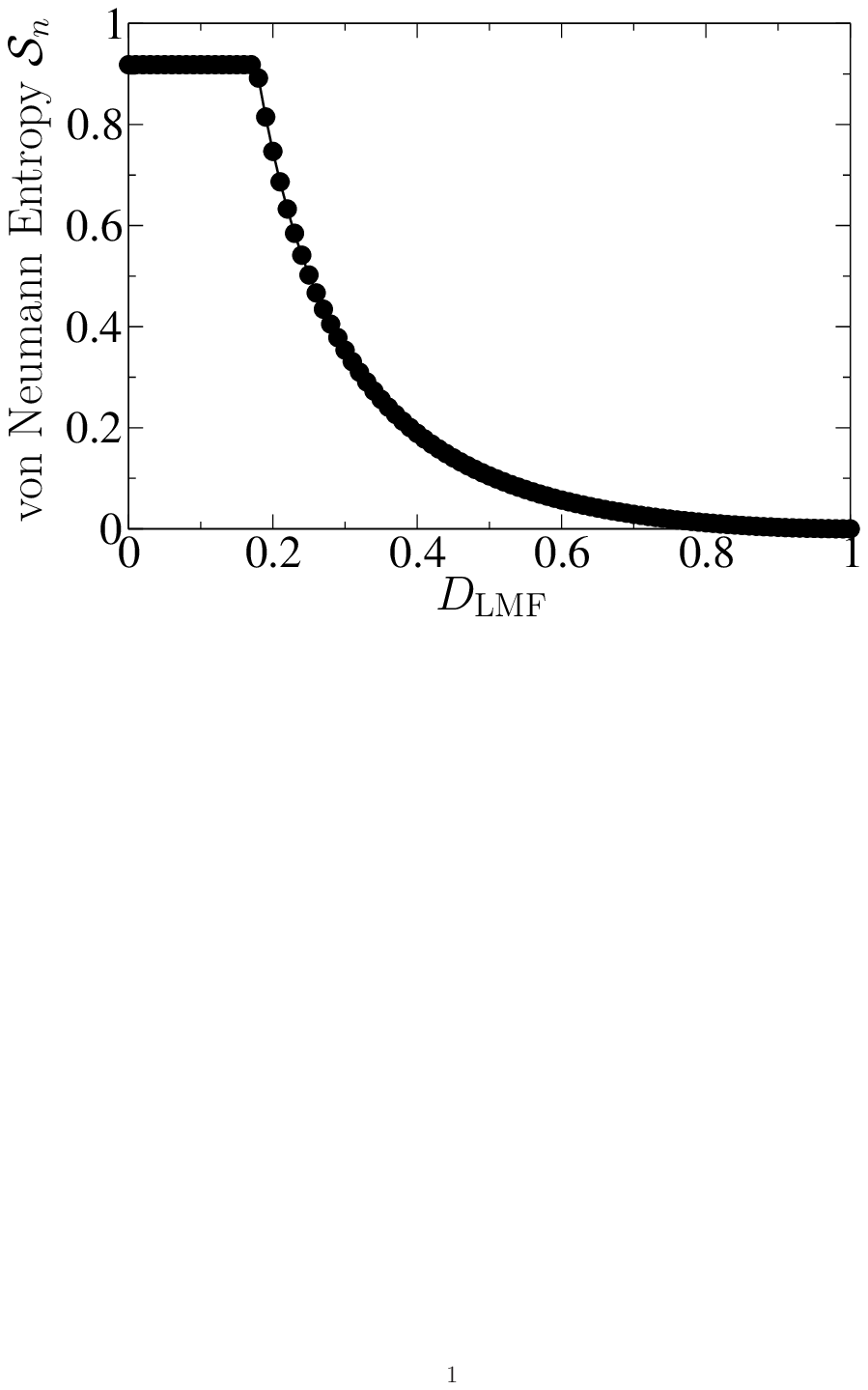}
  \end{center}
  \caption{Magnetization, principal angle, chirality, and von Neumann entropy (as a measure of entanglement) as functions of the classicality parameter $D_{\mathrm{LMF}}$, for a constant magnetic field $B = 0.4$.}  
  \label{f:pic3} 
\end{figure}

Regarding the transitions between the individual ground states, these are continuous and are expressed by various parameters, such as the magnetizations 
\begin{subequations}
  \begin{eqnarray}  
    M_n =|\langle \psi |\vec{S}_n|\psi \rangle|\;,
  \end{eqnarray} 
  the principal angles between the spins 
  \begin{eqnarray}  
    \Omega = \mathrm{arccos}\!\left(\frac{1}{S^2}\langle \psi|\vec{S}_n |\psi \rangle \cdot \langle \psi |\vec{S}_m |\psi \rangle\right)\;,
  \end{eqnarray} 
\end{subequations} 
the chirality $\chi$, as defined in Eq.~(\ref{ChiralityDef}), and the von Neumann entropies ${\cal S}_n$ (see the appendix). Please note that chirality applies only to non-coplanar spin structures. Coplanar spin structures, such as the coplanar $120^\circ$ texture, have zero chirality according to the definition. In the two equations above, $|\psi\rangle$ stands for one of the ground states $|\psi_1\rangle$, $|120^\circ(D_{\mathrm{LMF}},B)\rangle$, or $|\mathrm{PM}\rangle$. Because of the symmetric geometry of our system, $M_n$ and ${\cal S}_n$ are independent of the spin considered and $\Omega$ is independent of the pair of spins. 
  
To illustrate these transitions, we consider how the angles between the spin expectation values, the magnetization, the von Neumann entropy, and the chirality change along the curves $B = \mathrm{constant}$ or $D_{\mathrm{LMF}} = \mathrm{constant}$.

Figures \ref{f:pic2} and \ref{f:pic3} show these transitions for two different constant values of the external magnetic field. More specifically, Fig.~\ref{f:pic2} shows the magnetization, the principal angle between the spins, the chirality, and the von Neumann entropy for $B = 0$ (no magnetic field) as a function of the classicality $D_{\mathrm{LMF}}$, and  Fig.~\ref{f:pic3} shows the same plots for $B = 0.4$. 
  
One can see in Fig.~\ref{f:pic2} the transition among the ground states $|120^\circ(D_{\mathrm{LMF}},B) \rangle$ (which depend on the classicality) of the trimer without the influence of the external magnetic field. The ground states are always of the type $|120^\circ(D_{\mathrm{LMF}},B) \rangle$ independently of the classicality $D_{\mathrm{LMF}}$.  
At $B = 0$, the principal angle remains unchanged, which means that the spins are oriented in the $XY$-plane for all $D_{\mathrm{LMF}}$. Thus, the trimer exhibits no chirality in this case. At the same time, it can be seen that with increasing classicality, the entanglement, i.e., the value of the von Neumann entropy, decreases, reaching zero in the semiclassical limit. A decrease in entanglement simultaneously means an increase in magnetization towards the maximum value $M_n = 0.5$.

When the magnetic field $B$ is switched on, and below the threshold, $B < B_C$, we have a transition from the ground state $|\psi_1\rangle$, which does not depend on $B$ or $D_{\mathrm{LMF}}$, to a ground state $|120^\circ(D_{\mathrm{LMF}},B)\rangle$, see  Fig.~\ref{f:pic3}.
The state $|\psi_1\rangle$ exhibits the highest entanglement, chirality, and has a principal angle of zero (the spins are all oriented in the $z$-direction).

Entanglement decreases with increasing classicality in the ground states $|120^\circ(D_{\mathrm{LMF}},B)\rangle$, leading to an increase in magnetization. Likewise, a decrease in chirality can be observed, while the principal angle relatively quickly assumes a constant value. This, also considered in the context of the parallel orientation of the spins in $|\psi_1\rangle$, means that the enhanced chirality observed for $|\psi_1\rangle$ arises from genuinely quantum-mechanical three-spin correlations, Eq.~(\ref{ChiralityDef}), and not from a particular geometric spin arrangement, and vanishes as the system approaches the classical $|120^\circ(D_{\mathrm{LMF}},B)\rangle$ state. As the system becomes more classical, the quantum-mechanical three-spin correlations underlying the chirality are gradually replaced by effectively classical correlations between local spin expectation values, for which such higher-order correlators factorize. In this classical limit, the chirality is bounded by the value $\chi = 1/8$ for $S = 1/2$.

\begin{figure}[ht]
  \begin{center}
    \includegraphics*[width=4.0cm,bb = 70 450 510 770]{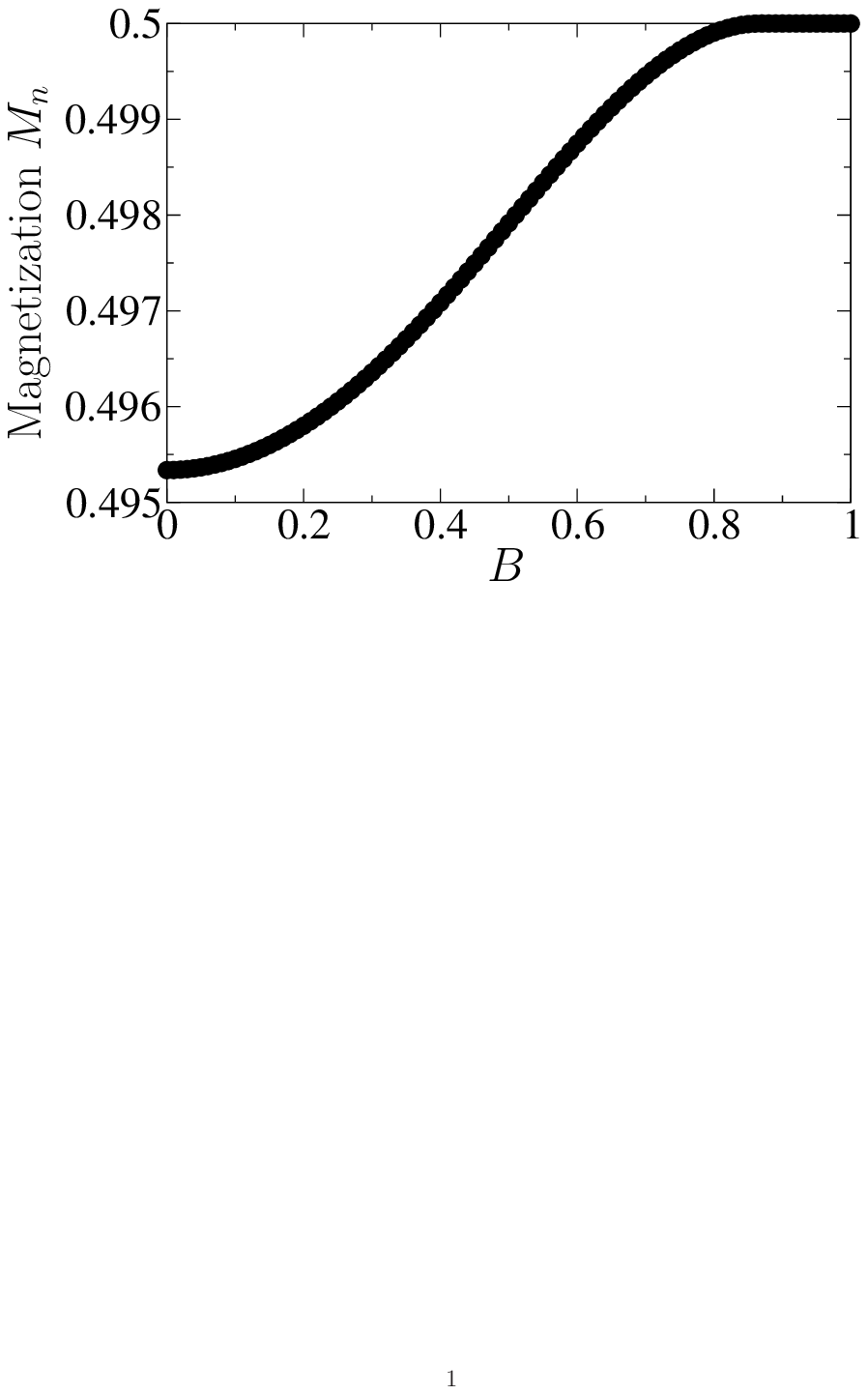}
    \includegraphics*[width=4.0cm,bb = 70 450 510 770]{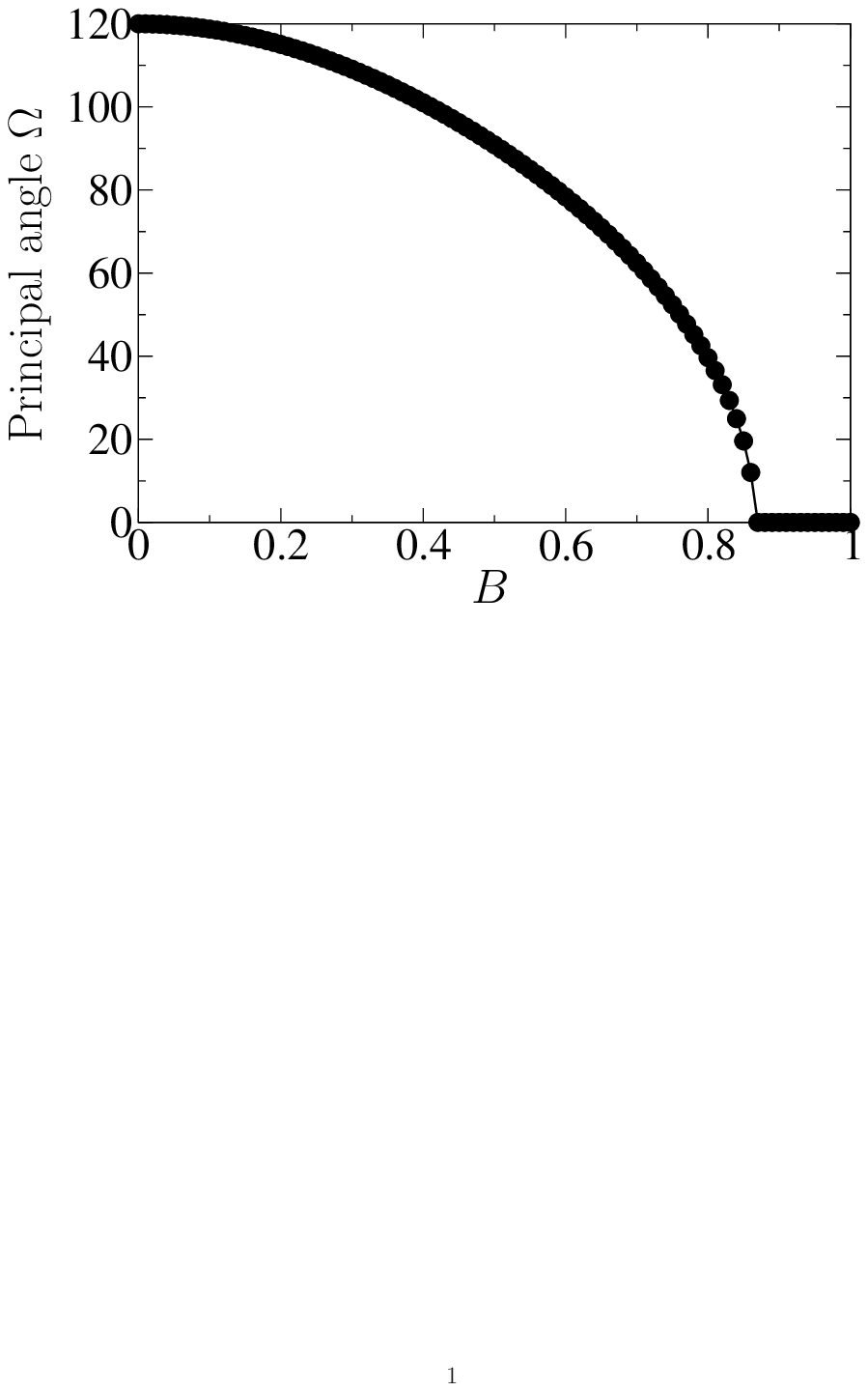}
    \includegraphics*[width=4.0cm,bb = 70 450 510 770]{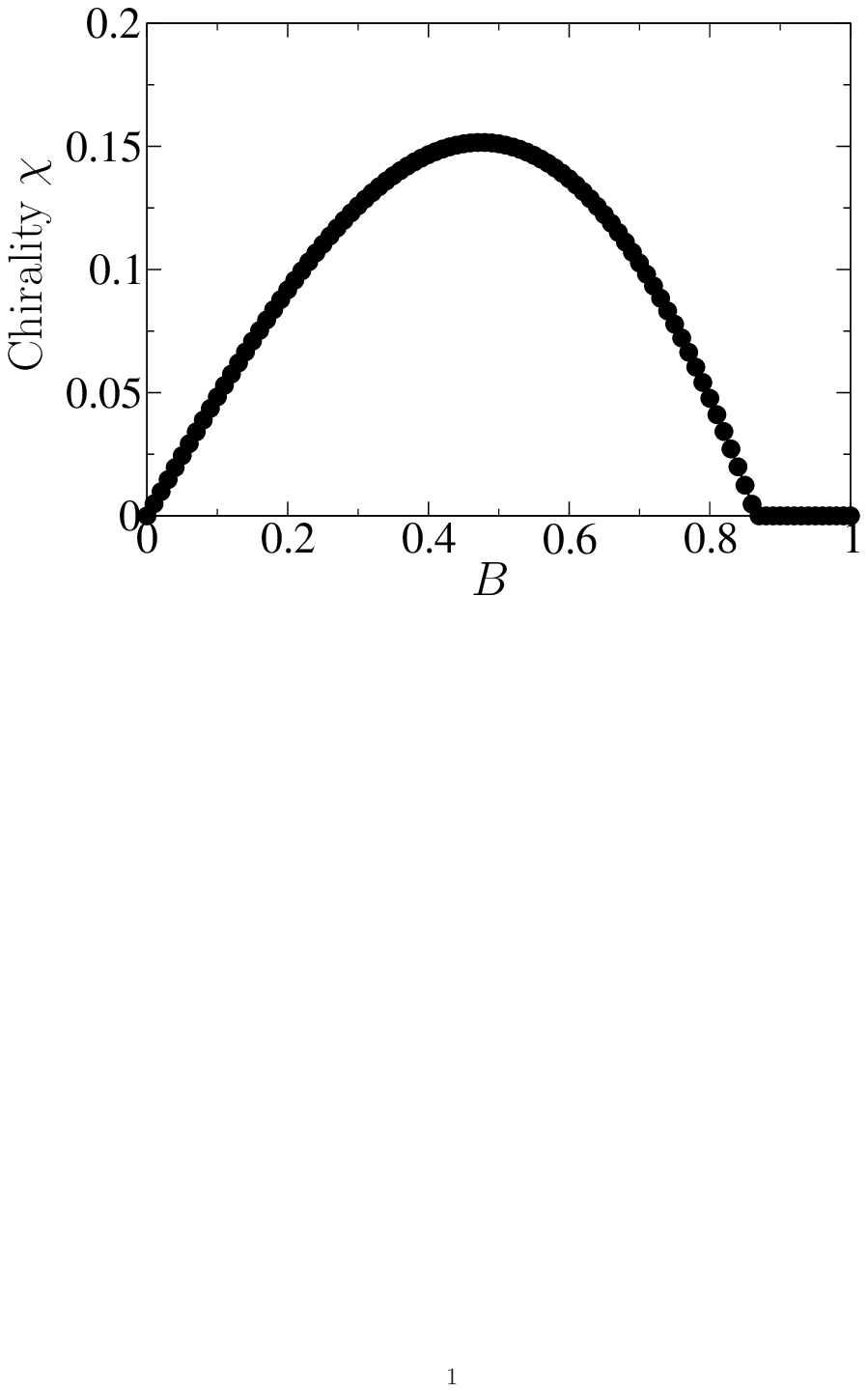}
    \includegraphics*[width=4.0cm,bb = 70 450 510 770]{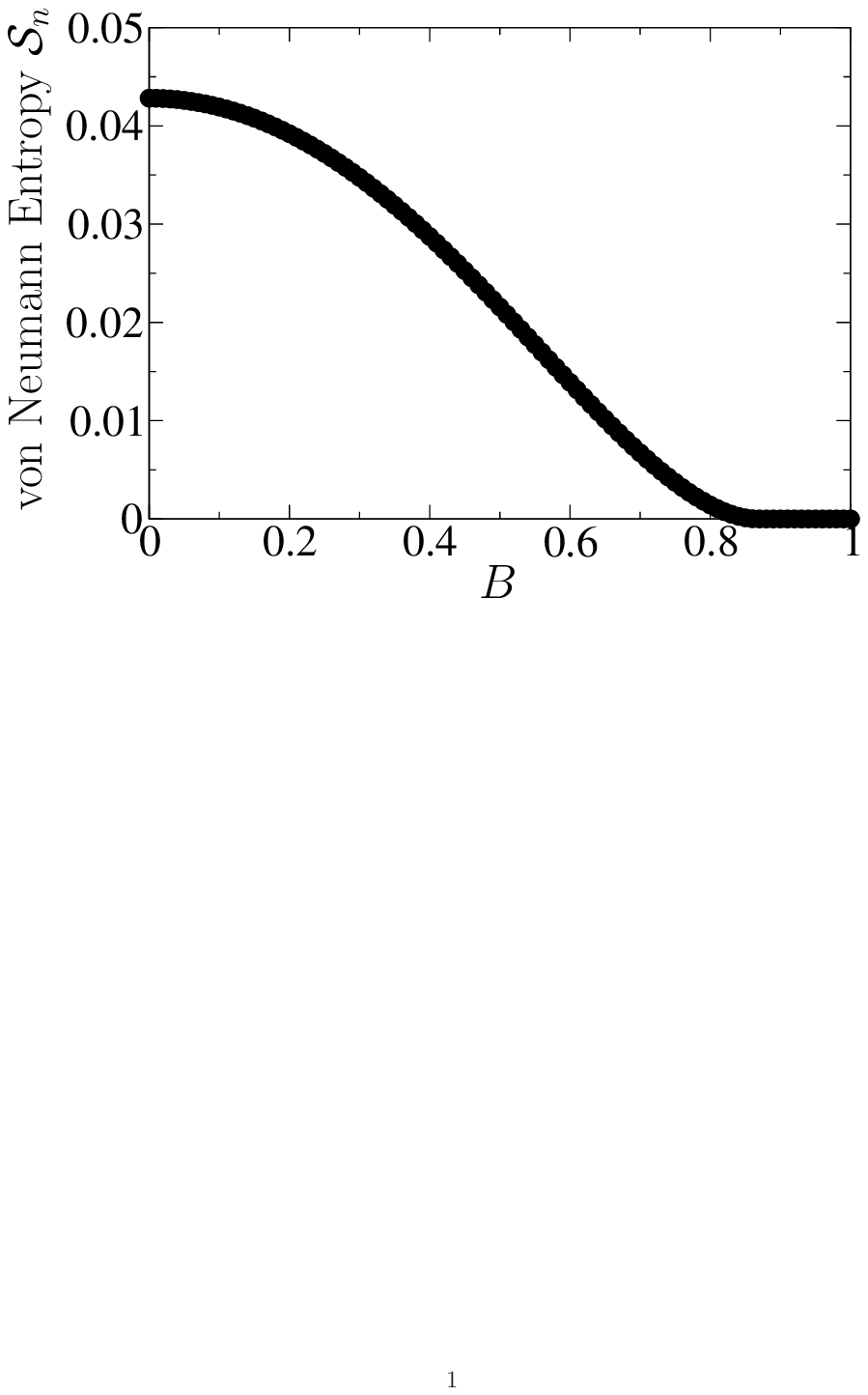}
  \end{center}
  \caption{Magnetization, principal angle, chirality, and von Neumann entropy (as a measure of entanglement) as functions of the external magnetic field $B$, at constant classicality $D_{\mathrm{LMF}} = 0.7$.}  
  \label{f:pic4}
\end{figure}

\begin{figure}[ht]
  \begin{center}
    \includegraphics*[width=4.0cm,bb = 70 450 510 770]{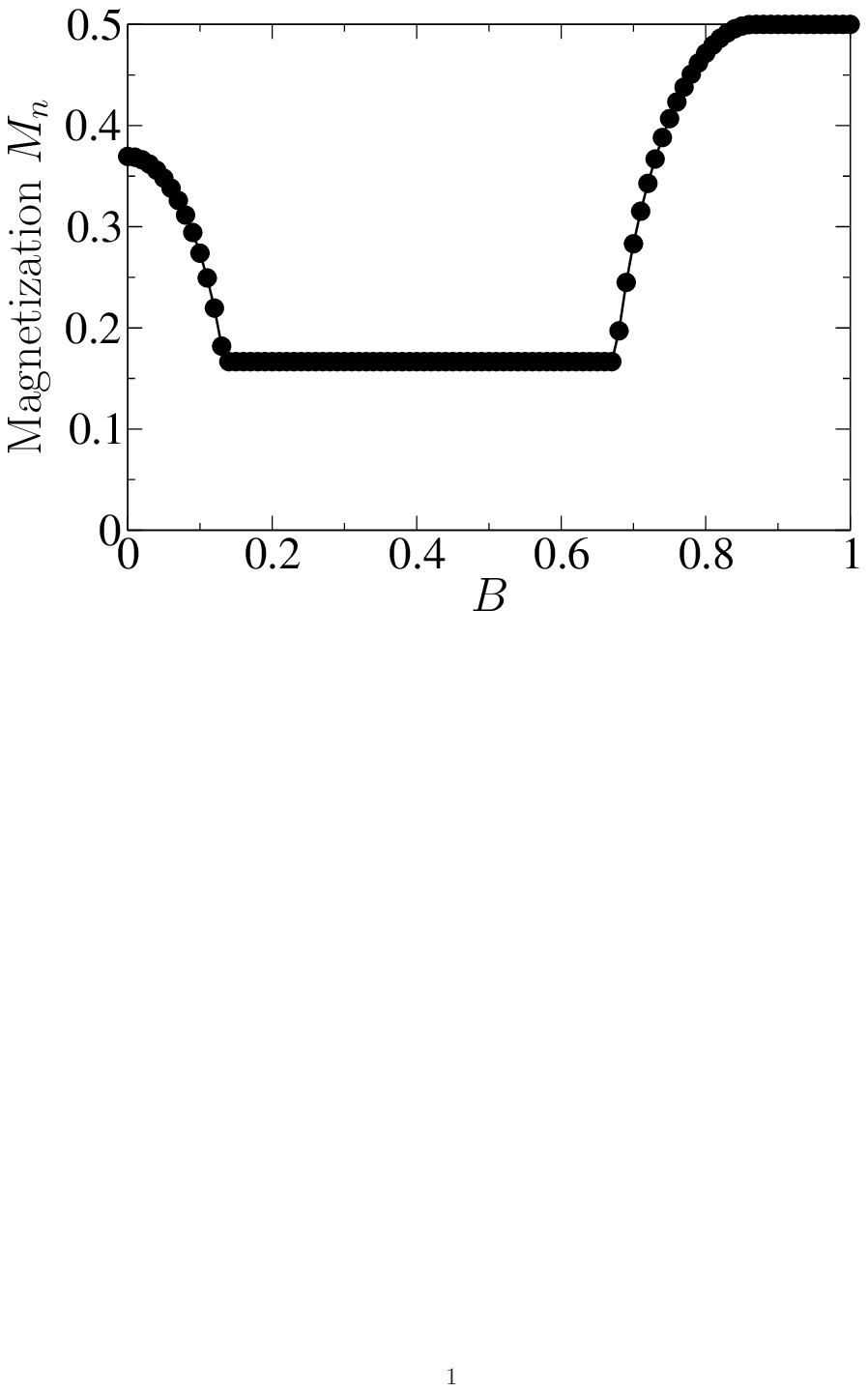}
    \includegraphics*[width=4.0cm,bb = 70 450 510 770]{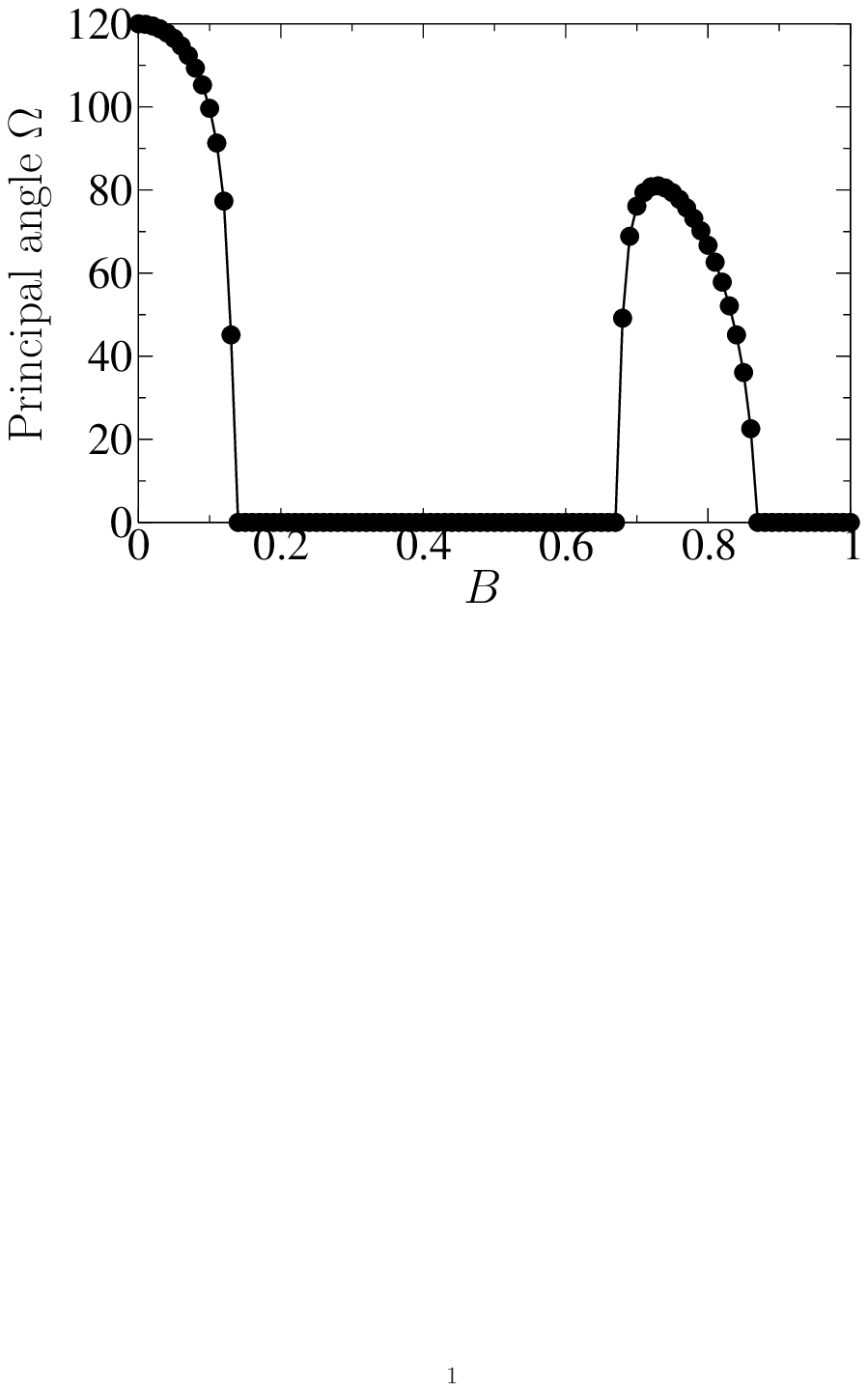}
    \includegraphics*[width=4.0cm,bb = 70 450 510 770]{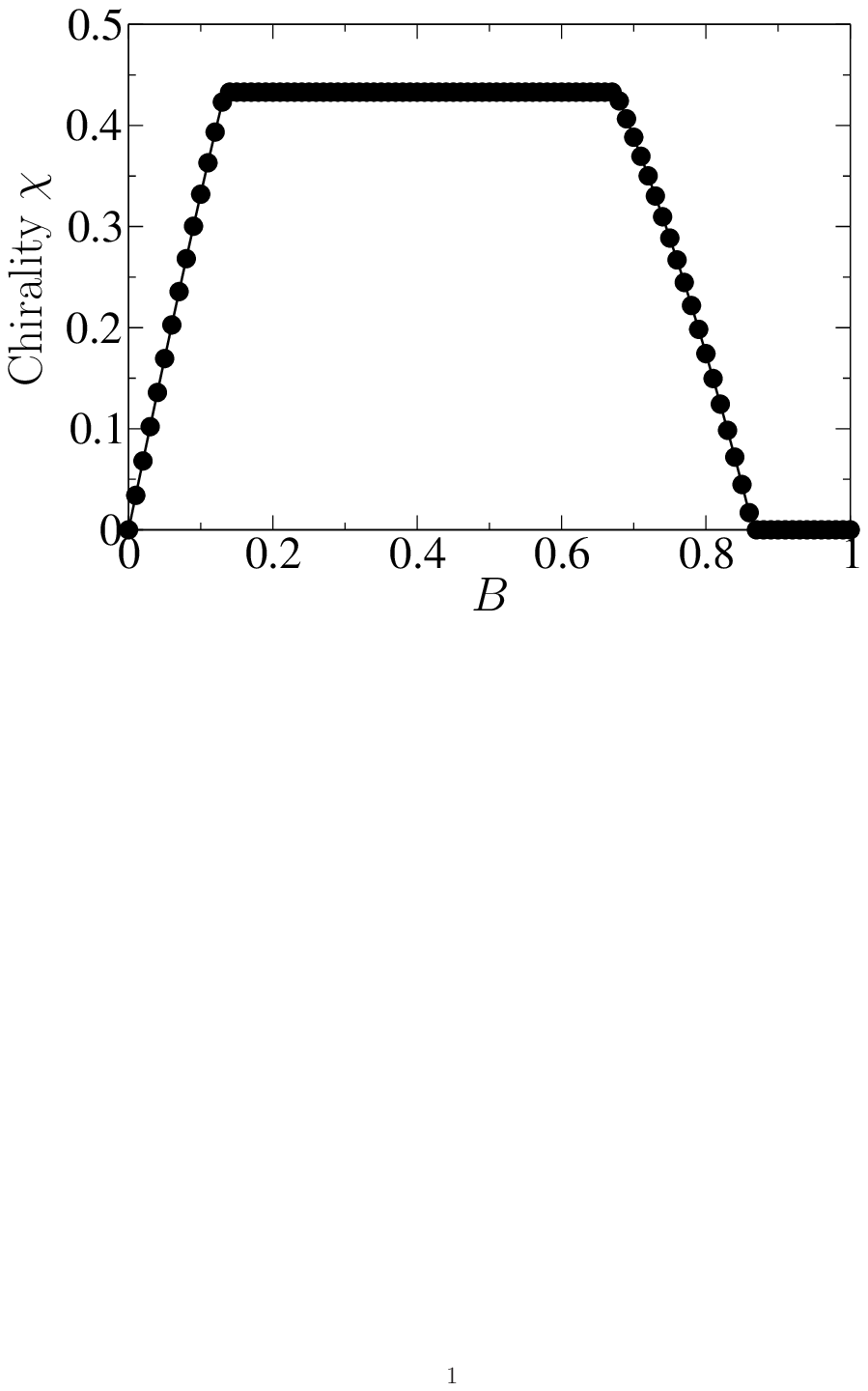}
    \includegraphics*[width=4.0cm,bb = 70 450 510 770]{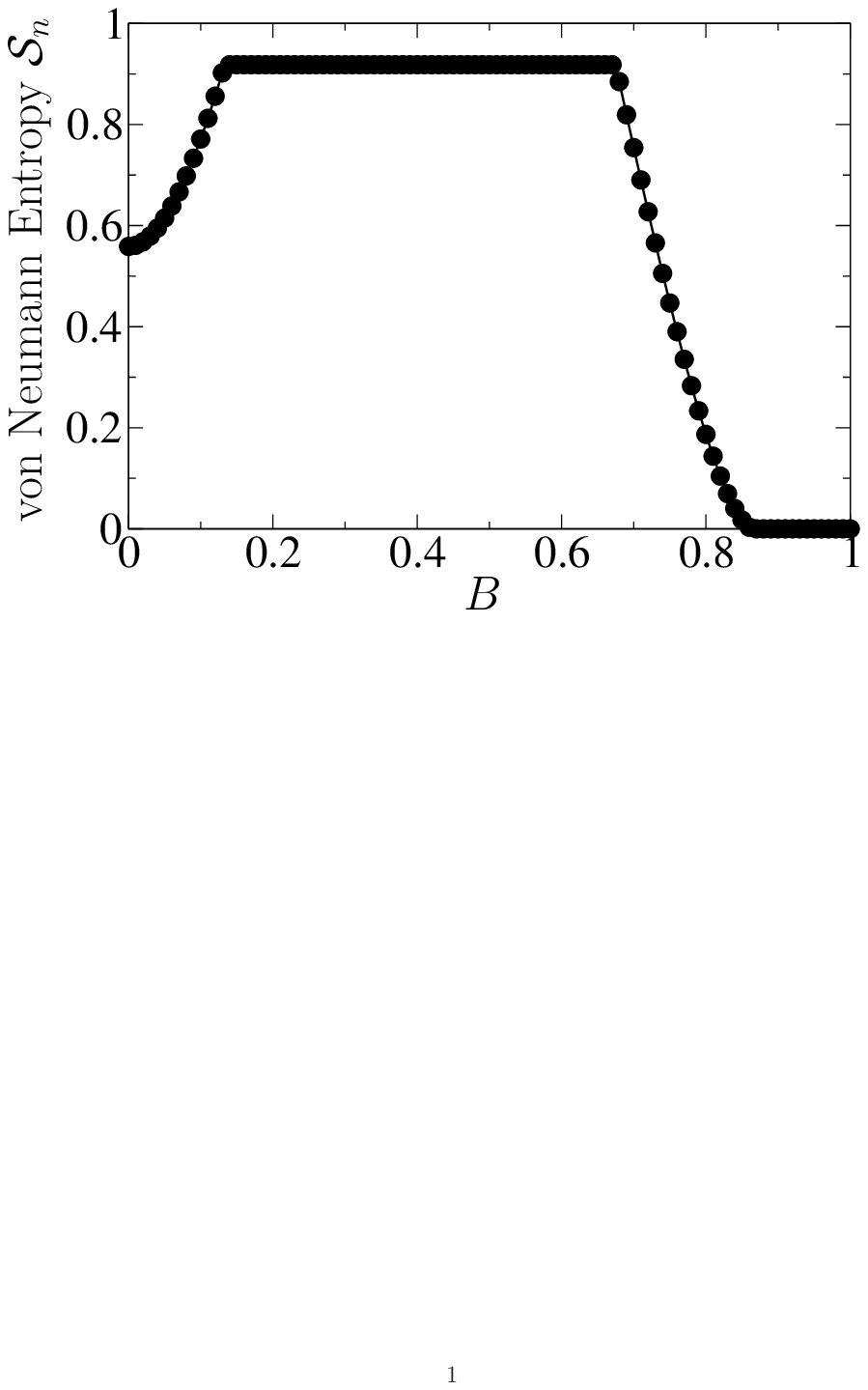}
  \end{center}
  \caption{Magnetization, principal angle, chirality, and von Neumann entropy (as a measure of entanglement) as functions of the external magnetic field $B$, at constant classicality $D_{\mathrm{LMF}} = 0.1$.}  
  \label{f:pic5}
\end{figure}

Curves with constant classicality and increasing external magnetic field describe the spin-flip transition toward the paramagnetic quantum state $|\mathrm{PM}\rangle$. Figs.~\ref{f:pic4} and \ref{f:pic5} show such transitions. Fig.~\ref{f:pic4} shows the magnetization, principal angle, chirality, and von Neumann entropy at $D_{\mathrm{LMF}} = 0.7$ as functions of the external magnetic field $B$. Here, a pure spin-flip transition can be observed. The principal angle changes continuously from $\Omega = 120^\circ$ to $0^\circ$. This spin-flip transition is caused by the external magnetic field. At the same time, the chirality changes, with the maximum occurring at $\Omega = 90^\circ$. 
The decrease in entanglement and the increase in magnetization with increasing magnetic field are universal. In this case, the magnetization increases steadily with the magnetic field until the saturation limit of $M_n = 1/2$. Observe that for $B=0$ the magnetization is close to $M_n = 1/2$ because we are close to the classical limit $D_{\mathrm{LMF}}=1$.  

Fig.~\ref{f:pic5} shows the same functions as in the previous figure but for $D_{\mathrm{LMF}} = 0.1$. It shows the different behavior of $|\psi_1\rangle$ and the ground states $|120^\circ(D_{\mathrm{LMF}},B) \rangle$. The ground state $|\psi_1\rangle$ is independent of $B$ or $D_{\mathrm{LMF}}$. In the quantum states $|120^\circ(D_{\mathrm{LMF}},B) \rangle$, however, there is a dependence on both $B$ and $D_{\mathrm{LMF}}$, and the spin-flip transition is again observed with decreasing principal angle and increasing magnetization. At the same time, entanglement decreases with increasing magnetic field, and chirality decreases from $B\approx 0.67$, which is the value of the magnetic field for which there is a transition from the ground state $|\psi_1\rangle$ to a ground state $|120^\circ(D_{\mathrm{LMF}},B) \rangle$.  

\section{Spin dynamics caused by the chirality}
\begin{figure*}[ht]
  \begin{center}
    \includegraphics*[width=7.0cm,bb = 70 430 510 770]{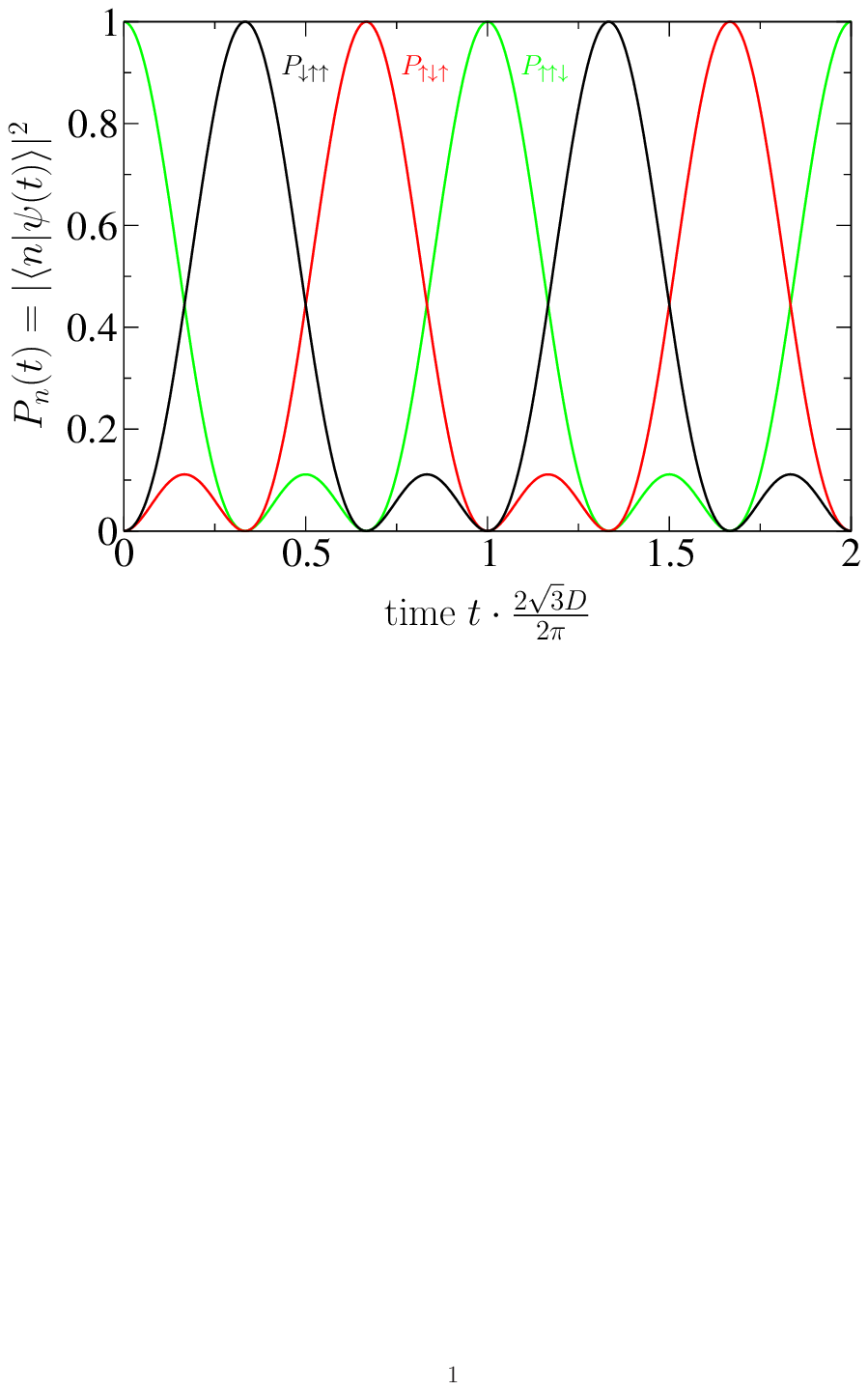}
    \includegraphics*[width=7.0cm,bb = 70 430 510 770]{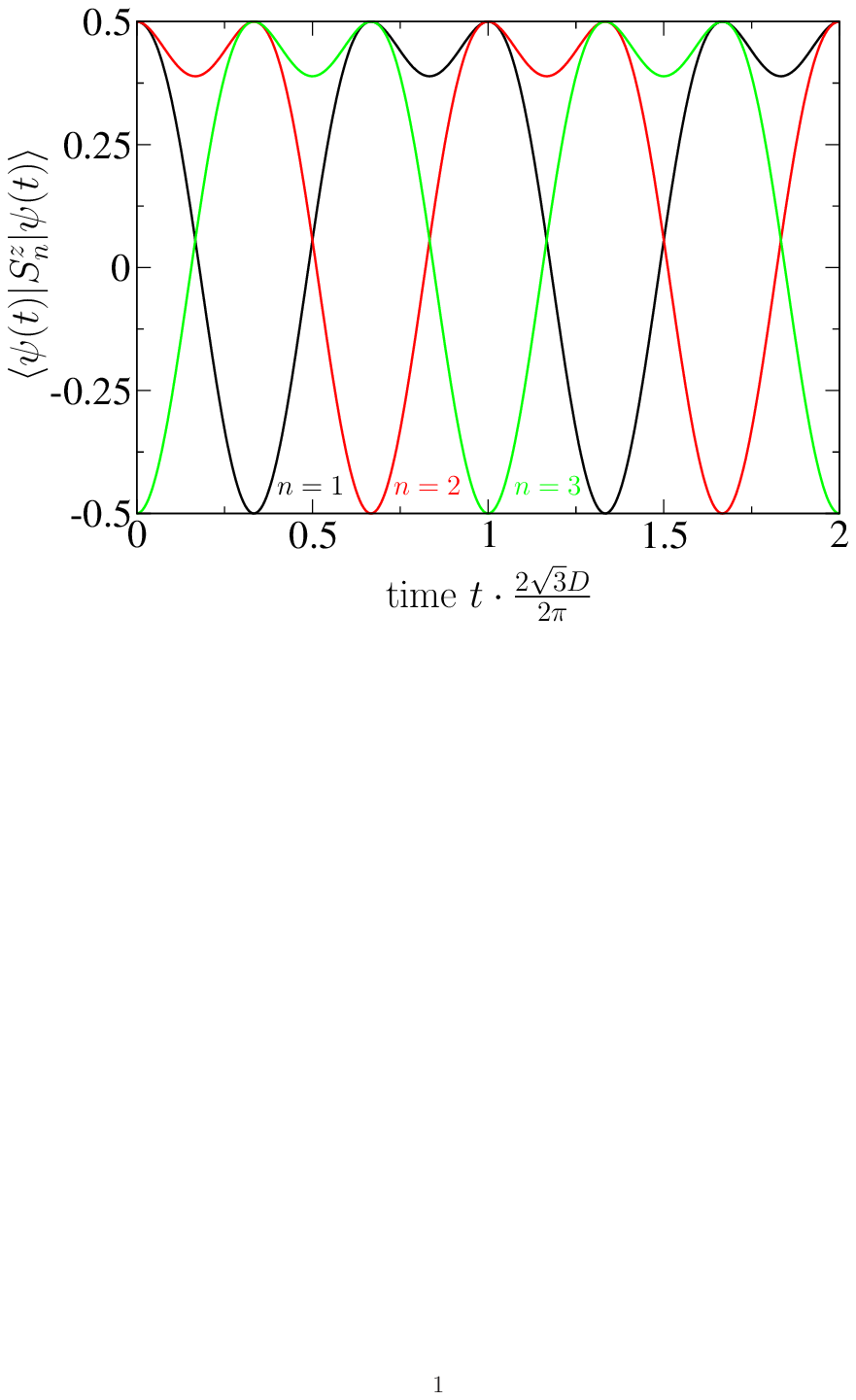}
  \end{center}
  \caption{(color online) Chiral spin dynamics: The probabilities $P_n(t) = |\langle n | \psi(t)\rangle|^2 = |a_n(t)|^2$ of finding the basis states $|n\rangle$ of $|\psi(t)\rangle$, Eq.~(\ref{3erSolution}), as well as the spin expectation values $\langle \psi(t)|\hat{S}_n^z | \psi(t) \rangle$ as a function of time $t$.}  
  \label{f:pic6}
\end{figure*}
The publications by Da-Wei Wang \emph{et al.} and Shuyue Wang \emph{et al.} demonstrate that the model system described in this publication (trimer with Dzyaloshinsky-Moriya interaction) can be realized experimentally \cite{wangNATUREPhys19,wangPRA24}. Da-Wei Wang \emph{et al.} and Shuyue Wang \emph{et al.} presented spin dynamics in which a flipped spin, respectively, pseudo-spin periodically rotates through the trimer.  This trimer was then used as a basis element for larger spin clusters, and periodic dynamics were also shown in these systems. The spin dynamics arising from the chirality are due to the Dzyaloshinsky-Moriya interaction. In this section, we will describe the dynamics analytically. Furthermore, we will study how the dynamics change during the quantum-classical transition. The spin dynamics is described by using the Hamilton operator $\hat{\mathrm{H}}$, Eq.~(\ref{HamTotal}), and by solving the modified Gisin-Schrödinger equation (\ref{eqGisin}), without damping, $\alpha = 0$, analytically, by calculating the eigensystem of $\hat{\mathrm{H}}$. With the initial quantum state: $|\psi(t=0)\rangle = |\!\uparrow\uparrow \downarrow\rangle$, we obtain the following solution:
  \begin{eqnarray} \label{3erSolution}
    |\psi(t)\rangle = e^{iBt}\Big[
      a_1(t) |\!\downarrow\uparrow\uparrow \rangle  +
      a_2(t) |\!\uparrow\downarrow\uparrow \rangle +
      a_3(t) |\!\uparrow\uparrow\downarrow \rangle \Big]\; 
  \end{eqnarray}
  with
  \begin{eqnarray}
    a_1(t) &=&
    \frac{1}{3}\left[1+\sqrt{3}\sin(2\sqrt{3}Dt)-\cos(2\sqrt{3}Dt)\right]
    \;, \label{3erSolutionB} \nonumber \\
    a_2(t) &=&
    \frac{1}{3}\left[1-\sqrt{3}\sin(2\sqrt{3}Dt)-\cos(2\sqrt{3}Dt)\right]
    \;, \label{3erSolutionC} \nonumber \\
    a_3(t) &=& \frac{1}{3}\left[1+2\cos(2\sqrt{3}Dt)\right]
    \;. \nonumber 
    \end{eqnarray}

Two aspects of this result are striking. First, the magnetic field does not influence the dynamics but merely enters as a global phase. 
Second, only the quantum component $D$ of the Dzyaloshinsky-Moriya interaction governs the coherent dynamics. The local mean-field part $D_{\mathrm{LMF}}$ does not contribute. In other words, the spin dynamics is purely quantum mechanical; there is no classical analog.

As already mentioned in Section \ref{sect_theo}, the coupling constants in the Dzyaloshinsky-Moriya interaction are constrained to satisfy: $D+D_{\mathrm{LMF}} = 1$. With increasing classicality, the frequency $\omega = 2\sqrt{3}\,D$ in the coefficients $a_n(t)$ decreases and becomes zero at $D_{\mathrm{LMF}}=1$. The dynamics thus cease in the semiclassical limit, as the coefficients no longer depend on time.

Using the quantum state $|\psi(t)\rangle$ from Eq.~(\ref{3erSolution}), one can easily calculate the spin expectation values $\langle \hat{S}_n^\eta \rangle = \langle \psi(t)| \hat{S}_n^\eta |\psi(t) \rangle$, for $n = 1,2,3$ and $\eta = x,y,z$. The components of the spin expectation values in the $XY$-plane are zero, and the $z$-components are:
\begin{subequations}
  \begin{eqnarray}
    \langle \hat{S}_1^z \rangle &=& 
    \frac{1}{6}\!+\!\frac{1}{9}\!\left(\cos(4\sqrt{3}Dt) \!+\! \sqrt{3}\sin(4\sqrt{3}Dt) \right. \nonumber \\
    && \left. + 2 \cos(2\sqrt{3}Dt) \!-\! 2 \sqrt{3} \sin(2\sqrt{3}Dt)\right) 
    \;,  \\ 
    \langle \hat{S}_2^z \rangle &=& \frac{1}{6} \!+\!\frac{1}{9}\!\left(\cos(4\sqrt{3}Dt) \!-\! \sqrt{3}
    \sin(4\sqrt{3}Dt) \right. \nonumber \\
    && \left. + 2 \cos(2\sqrt{3}Dt) \!+\! 2\sqrt{3} \sin(2\sqrt{3}Dt)\right)  \;,  \\
    \langle \hat{S}_3^z \rangle &=& \frac{1}{6}  \!-\! \frac{1}{9}\!\left(2
    \cos(4\sqrt{3}Dt) \!+\! 4 \cos(2\sqrt{3}Dt) \right)  \;. 
  \end{eqnarray}
\end{subequations}
If we denote the basis states as $|1\rangle = |\!\downarrow\uparrow\uparrow\rangle$, $|2\rangle  = |\!\uparrow\downarrow\uparrow\rangle$, $|3\rangle = |\!\uparrow\uparrow\downarrow\rangle$, then the probabilities of finding $|\psi(t)\rangle$ in one of these basis states are $P_n(t) = |\langle n | \psi(t)\rangle|^2 = |a_n(t)|^2$.

Figure \ref{f:pic6} shows these probabilities, as well as the spin expectation values $\langle \psi(t)| \hat{S}_n^z |\psi(t) \rangle$ for $n = 1, 2, 3$, of the chiral spin dynamics. The dynamics are longitudinal and, apart from the presence of thermal and quantum fluctuations in the experiment of Da-Wei Wang \emph{et al.} and Shuyue Wang \emph{et al.}, are consistent with the results reported in \cite{wangNATUREPhys19,wangPRA24}. In this process, the spin oriented opposite to the other two, rotates cyclically through the spins of the trimer. The orientation and rotation frequency are determined by the Dzyaloshinsky–Moriya interaction parameter $D$.

Finally, the dynamics of the entanglement between the three spins will be considered.
\begin{figure}[ht]
  \begin{center}
    \includegraphics*[bb =75 440 515 765,width=7cm]{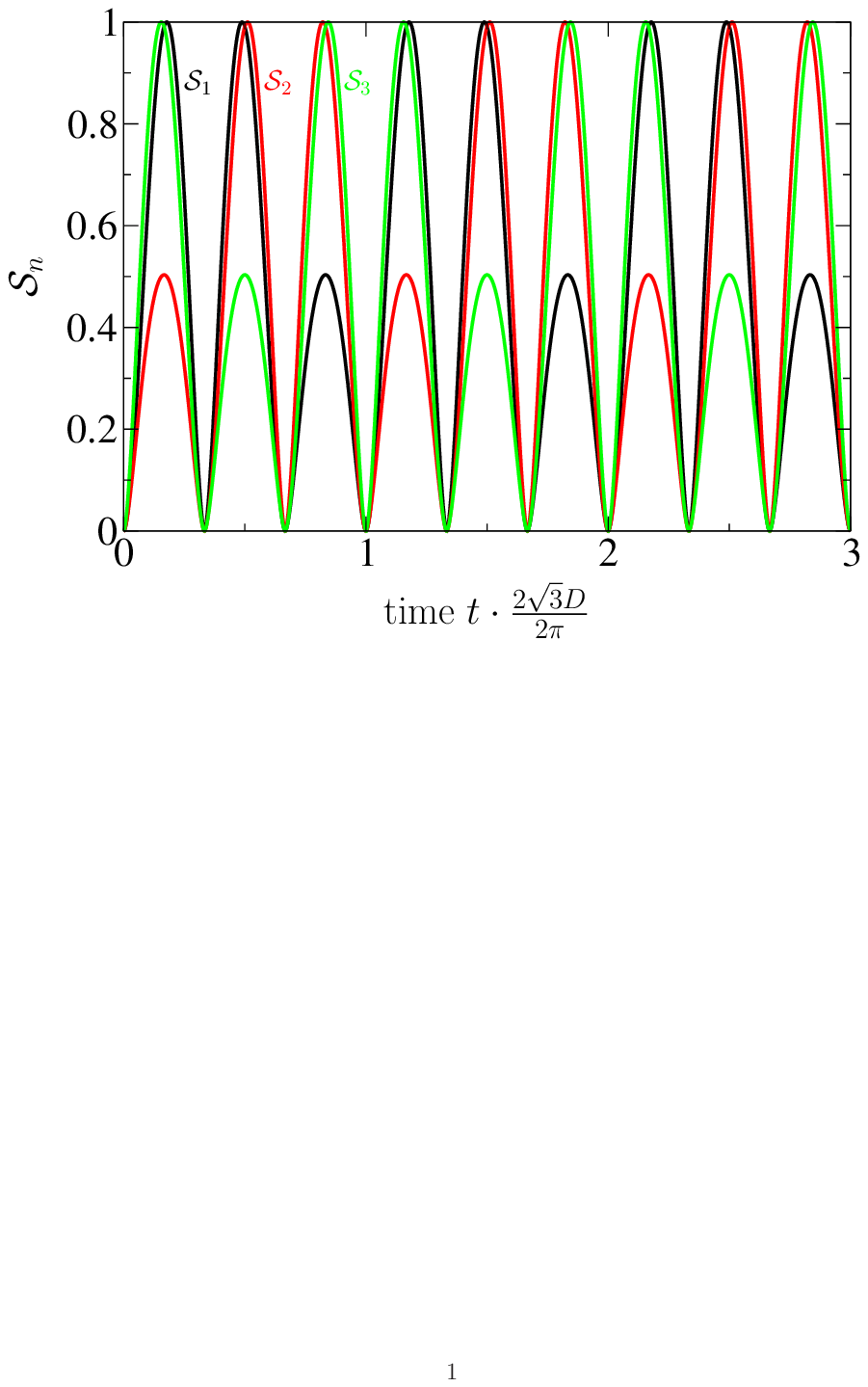}
  \end{center}
  \caption{(color online) Chiral spin dynamics: The von Neumann entropies ${\cal S}_n$ as a function of time $t$.}  
  \label{f:pic7}
\end{figure}
The details for calculating the von Neumann entropies are given in the appendix. For our system, the von Neumann entropy contribution of each spin is given by
\begin{equation}
  {\cal S}_n(t) = -\big(1-|a_n|^2\big)\,\log_2\!\big(1-|a_n|^2\big)
   - |a_n|^2 \log_2 |a_n|^2 \;.
\end{equation}
Figure \ref{f:pic7} shows the von Neumann entropies ${\cal S}_n$ as functions of time $t$. It exhibits the same periodic pattern as the spin expectation values and the probabilities, $P_n(t)$. Whenever the quantum state $|\psi(t)\rangle$ is identical to one of the basis states, the von Neumann entropies ${\cal S}_n$ are zero, otherwise, two spins are always thoroughly entangled. These are the two spins between which the rotated spin is exchanged. The remaining third spin exhibits reduced entanglement. Here, the value of the von Neumann entropy for the two entirely entangled spins achieves the maximum value ${\cal S}_n = 1$. The third spin, however, has reduced entanglement with ${\cal S}_n = 1/2$.

\section{Summary}
In this work, the quantum-classical transition of a trimer with Dzyaloshinsky-Moriya interaction was investigated. The model system is described by a Hamilton operator, which, in addition to the pure quantum mechanical description using interacting spin operators, also contains a semiclassical description given by the interaction of each spin with the expectation values of the other spins (local mean-fields). The two contributions in the Dzyaloshinsky-Moriya term are weighted so that the interaction parameters satisfy: $D + D_{\mathrm{LMF}} = 1$. The result of the calculation, taking into account an external magnetic field in the same direction as $\vec{D}$ and $\vec{D}_{\mathrm{LMF}}$, was then plotted in a ground state diagram. This ground state diagram contains the ground states $|\mathrm{PM}\rangle$, $|\psi_1\rangle$, and $|120^\circ(D_{\mathrm{LMF}},B)\rangle$. The quantum state $|\mathrm{PM}\rangle$ is a product state corresponding to the paramagnetic configuration of the trimer in which the Dzyaloshinsky-Moriya interaction has no effect. This state appears when the external magnetic field exceeds a critical value $B_C$. The quantum state $|\psi_1\rangle$ is characterized by being the ``most quantum mechanical'' configuration, in the sense that it exhibits a stable entanglement, and remains unaffected by both the external magnetic field and classicality. In contrast, the ground states $|120^\circ(D_{\mathrm{LMF}},B)\rangle$, depend on both the external magnetic field (through spin-flip transitions) and classicality. These states occupy the largest area of the ground state diagram, and in the limit $D_{\mathrm{LMF}}=1$,
these states coincide with the classical trimer with Dzyaloshinsky-Moriya interaction. 

In the second part of the work we analyzed the chiral spin dynamics proposed by
Da-Wei Wang \emph{et al.}. Starting from a non-eigenstate with one flipped
spin, we solved the modified Gisin-Schrödinger equation analytically and
obtained explicit expressions for the time evolution of state probabilities,
spin expectation values, and single-spin von Neumann entropies. The results
show a periodic cyclic rotation of the flipped spin through the trimer, a
purely quantum phenomenon absent in the semiclassical limit. These findings
provide a minimal analytical model that connects entanglement, chirality, and
quantum coherence in frustrated spin clusters, and may serve as a theoretical
framework for future experimental realizations.\\

\appendix
\section*{Appendix}
\subsection{Relaxation dynamics towards eigenstates of $\hat{\mathrm{H}}$}
For an initial state $|\psi(t=0)\rangle$ that is not an eigenstate of the Hamiltonian $\hat{\mathrm{H}}$, the modified Gisin–Schr\"odinger equation with $\alpha > 0$, induces a non-unitary dynamics that causes the state to relax toward an eigenstate of $\hat{\mathrm{H}}$, corresponding to a lower energy.
To see this, one considers the time derivative of the energy expectation value:
\begin{eqnarray}
    \frac{\mathrm{d}E(t)}{\mathrm{d}t} = \left(\frac{\mathrm{d}}{\mathrm{d}t}\langle \psi(t)|\right)\!\hat{\mathrm{H}}|\psi(t)\rangle + \langle \psi(t)|\hat{\mathrm{H}}\!\left(\frac{\mathrm{d}}{\mathrm{d}t}|\psi(t)\rangle\right) \;.
\end{eqnarray}
Inserting the modified Gisin-Schr\"odinger equation (\ref{eqGisin}) and the corresponding complex-conjugate equation leads to:
\begin{eqnarray} \label{eq:EchangeSuppl}
    \frac{\mathrm{d}E(t)}{\mathrm{d}t} = - \frac{2\alpha}{(1+\alpha^2)\hbar}\left(\langle \hat{\mathrm{H}}^2 \rangle - \langle \hat{\mathrm{H}} \rangle^2\right) \leq 0\;.
\end{eqnarray}
The interpretation of this formula is as follows: the energy decreases monotonically, and the change stops when the variation $\mathrm{Var}(\hat{\mathrm{H}}) = \langle \hat{\mathrm{H}}^2 \rangle - \langle \hat{\mathrm{H}} \rangle^2$ vanishes. This is precisely the case when $|\psi(t)\rangle$ is an eigenstate of $\hat{\mathrm{H}}$. Thus, $E(t)$ is a Lyapunov function of the dynamics, with the fixed points $\mathrm{Var}(\hat{\mathrm{H}}) = 0$.

\subsection{Connection between the modified Gisin-Schr\"odinger and the LLG equation}
For a single spin-$1/2$: $\vec{S} = \frac{\hbar}{2} \,\vec{\sigma}$, with gyromagnetic ratio $\gamma$ and Hamiltonian
\begin{eqnarray}
  \hat{\mathrm{H}} = -\gamma \frac{\hbar}{2}\, \vec{B} \cdot \vec{\sigma} \;,
\end{eqnarray}
the time derivative of the expectation value $\langle \hat{\sigma}_r \rangle = \langle \psi(t)|\hat{\sigma}_r|\psi(t)\rangle$, $r \in \{x,y,z\}$, is given by 
\begin{equation}
\frac{\mathrm{d}\langle\hat{\sigma}_r\rangle}{\mathrm{d}t}
= \left(\frac{\mathrm{d}}{\mathrm{d}t}\langle\psi(t)|\right)\!\hat{\sigma}_r|\psi(t)\rangle
  + \langle\psi(t)|\hat{\sigma}_r\!\left(\frac{\mathrm{d}}{\mathrm{d}t}|\psi(t)\rangle\right).
\label{eq:sigma_derivative}
\end{equation}
Inserting the modified Gisin-Schr\"odinger equation, Eq.~\eqref{eqGisin} and its conjugate-complex version, and using the commutator and anti-commutator relations,
\begin{align}
[\hat{\sigma}_r,\hat{\sigma}_s] &= 2i\,\varepsilon_{rst}\,\hat{\sigma}_t, \label{eq:comm}\\
\{\hat{\sigma}_r,\hat{\sigma}_s\} &= 2\,\delta_{rs}\,\mathbf{1}, \label{eq:anticomm}
\end{align}
where $\mathbf{1}$ is the $2\times 2$ unit matrix, $\delta_{rs}$ is the Kronecker and $\varepsilon_{rst}$ the Levi--Civita symbol, one obtains
\begin{equation}
(1+\alpha^2)\frac{\mathrm{d}\langle\hat{\sigma}_r\rangle}{\mathrm{d}t}
= \gamma\,\Big(\langle \vec{\sigma} \rangle \times \vec{B}\Big)_r
  - \gamma \alpha\,\Big(\langle \vec{\sigma} \rangle \times (\langle \vec{\sigma} \rangle \times \vec{B})\Big)_r.
\label{eq:sigma_dynamics}
\end{equation}
We recover the Landau-Lifshitz-Gilbert equation with $\vec{m} = \langle \vec{\sigma} \rangle$, and $|\vec{m}| = 1$. 

In the case of $N$ spins, classical spin dynamics occur if the magnetic field $\vec{B}$ is a local mean-field, e.g. 
\begin{eqnarray}
  \hat{\mathrm{H}} = D_{\mathrm{LMF}} \sum_{n} \vec{z}\cdot \!\Big(
\vec{S}_n \times \langle \vec{S}_{n+1} \rangle \Big) = -\sum_{n} \vec{B}_n \cdot \vec{S}_n \;,
\end{eqnarray}
with 
\begin{equation}
\vec{B}_n = -\frac{\mathrm{d}\hat{\mathrm{H}}}{\mathrm{d}\vec{S}_n} = -D_{\mathrm{LMF}}\!\left(\begin{array}{c}
    \langle \hat{S}_{n+1}^y \rangle - \langle \hat{S}_{n-1}^y \rangle \\
    \langle \hat{S}_{n-1}^x \rangle - \langle \hat{S}_{n+1}^x \rangle \\
    0 \end{array} \right),
\label{eq:B_localDMI}
\end{equation}
and the initial quantum state is a product state:
\begin{eqnarray}
  |\psi\rangle = \bigotimes\limits_{n=1}^N |\psi_n\rangle \;, 
\end{eqnarray}  
with
\begin{eqnarray} 
  |\psi_n\rangle = \cos(\theta_n/2)|\!\uparrow\rangle + \exp(i\phi_n)\sin(\theta_n/2)|\!\downarrow\rangle\;.
\end{eqnarray} 
\subsection{von Neumann Entropy}
In the case of a pure quantum state, the density matrix $\rho$ is given by: 
\begin{eqnarray}
  \rho = |\psi\rangle\langle\psi| \;,
\end{eqnarray}  
where $|\psi\rangle$ and $\langle\psi|$ are respectively, the ket and bra of the given quantum state. 

The von Neumann entropy is calculated from the partial density matrix  $\rho_n$, which results from the density matrix $\rho$ by tracing over the space of two spins. More precisely, the partial density matrices are given by:
\begin{eqnarray}
  \rho_n = \mathrm{Tr}_{\{1,2,3\}\setminus\{n\}}(\rho)\,, \qquad n=1,2,3.
\end{eqnarray}
In particular, for $n=1$ (analogous for $n=2,3$) the entries of the partial density matrix are,
\begin{equation}
(\rho_1)_{\alpha\beta}
=
\sum_{s_2,s_3\in\{\uparrow,\downarrow\}}
\langle \alpha\,s_2 s_3 | \rho | \beta\,s_2 s_3\rangle,
\qquad \alpha,\beta\in\{\uparrow,\downarrow\}.
\end{equation}
If $|\psi(t)\rangle =
a_1(t)\,|{\downarrow\uparrow\uparrow}\rangle
+ a_2(t)\,|{\uparrow\downarrow\uparrow}\rangle
+ a_3(t)\,|{\uparrow\uparrow\downarrow}\rangle$,
 with
$|a_1|^2+|a_2|^2+|a_3|^2=1$, then the partial density matrices are,
\begin{eqnarray}
  \rho_n = \left(\!\begin{array}{cc} 1- |a_n|^2 & 0 \\ 0 &
    |a_n|^2 \end{array}\!\right) \;.
\end{eqnarray}
The von Neumann entropy of the spin $n$, which quantifies its entanglement with the other two spins, is then 
\begin{align}
    {\cal S}_n(t)
&= -\mathrm{Tr}\,\big[\rho_n(t)\,\log_2 \rho_n(t)\big] \notag   \\
&= -\big(1-|a_n|^2\big)\,\log_2\!\big(1-|a_n|^2\big)
   - |a_n|^2 \log_2 |a_n|^2.
\end{align}

\begin{acknowledgments}
R. Wieser acknowledges the financial support provided by
the Startup Foundation for Introducing Talent of NUIST (2018r043).  
\end{acknowledgments}

\bibliography{Cite}

\end{document}